\documentclass[final,authoryear,3p,times,twocolumn]{elsarticle}

\usepackage[latin2]{inputenc}
\usepackage{graphicx}

\usepackage{amssymb}
\usepackage{amsthm}
\usepackage{ulem}

\journal{Applied Radiation and Isotopes 107(2016)391}

\begin{document}

\begin{frontmatter}



\title{Investigation of activation cross sections of proton induced reactions on indium up to 70 MeV for practical applications}


\author[1]{F. T\'ark\'anyi}
\author[1]{F. Ditr\'oi\corref{*}}
\author[2]{A. Hermanne}
\author[1]{S. Tak\'acs}
\author[3]{M. Baba}
\author[3]{A.V. Ignatyuk}

\cortext[*]{Corresponding author: ditroi@atomki.hu}

\address[1]{Institute for Nuclear Research, Hungarian Academy of Sciences (ATOMKI),  Debrecen, Hungary}
\address[2]{Cyclotron Laboratory, Vrije Universiteit Brussel (VUB), Brussels, Belgium}
\address[2]{Cyclotron Radioisotope Center (CYRIC), Tohoku University, Sendai, Japan}

\begin{abstract}
Excitation functions were measured for production of the $^{113,111,110}$Sn,  $^{115m,114m,113m,112m,111g,110g}$In and $^{111m,109}$Cd radioisotopes by bombardment of In targets with proton  beams up to 70 MeV, some of them for the first time. The new results are compared with the earlier experimental data and with the theoretical data in the TENDL-2014 (Talys1.6 based) library. Thick target yields were deduced and application of the new data for production of medically relevant $^{110m}$In, $^{111g}$In, $^{113m}$In and $^{114m}$In, as well as applicability for thin layer activation (TLA) are  discussed.
\end{abstract}


\end{frontmatter}


\section{Introduction}
\label{1}
The aim of the present work was to measure more accurate and new cross sections for the production of the $^{113,111,110}$Sn,  $^{115m,114m,113m,112m,111g,110g}$In and $^{111m,109}$Cd radioisotopes. We also investigated the production capabilities  of   diagnostic and therapeutic radionuclides  $^{110m}$In, $^{111g}$In, $^{113m}$In  and  $^{114m}$In through the $^{nat}$In(p,x)$^{110}$Sn$\rightarrow$$^{110m}$In, $^{nat}$In(p,x)$^{111}$Sn$\rightarrow$$^{111g}$In, $^{nat}$In(p,xn)$^{113}$Sn$\rightarrow$$^{113m}$In and $^{nat}$In(p,x)$^{114m}$In reactions, in the frame of a systematic study of charged particle production routes of medical  radionuclides. 
The experimental data effectively contribute to practical applications via production of recommended cross sections in optimal energy ranges and derived yield data for production of the particular radioisotope for application purposes.
Another important result of the present study, the applicability of the new data in the field of wear measurements by using thin layer activation (TLA) is also discussed in the Application section.
The cross section data for production of some of these radioisotopes were investigated by us earlier in detail using different production routes: For production of $^{110}$Sn and for direct production of $^{110m}$In we  published  investigations in  \citep{Ali, Hermanne2010, Szelecsenyi1990, Szelecsenyi1991, Takacs2010, TF2011, TF2006, TF2007, TF2015, TF2004}.
Data for production of $^{111g}$In  were reported in  \citep{Hermanne2014b, Hermanne2014, Hermanne2006, Szelecsenyi1994, TF2011, TF2006, TF1994, TF2005b}.
For production of $^{113}$Sn and for direct production of $^{113m}$In we have already presented related experimental results on $^{nat}$Cd(p,xn)$^{113m}$In \citep{TF2005, TF2006}, $^{nat}$Cd(d,xn)$^{113}$In \citep{TF2007}, $^{114}$Cd(d,3n)$^{113m}$In \citep{TF2005b}, $^{nat}$Cd($^{3}$He,xn)$^{113}$Sn \citep{Szelecsenyi1991}, $^{nat}$Sn(p,x)$^{113}$Sn \citep{Hermanne2006}, $^{nat}$Cd($\alpha$,xn)$^{113}$Sn \citep{Hermanne2010} and recently  the $^{nat}$In(d,x)$^{113}$Sn reaction \citep{TF2011}. 
We have investigated the production routes of $^{114m}$In in  \citep{Hermanne2014b, Hermanne2010b, TF2011, TF2006, TF2007, TF2005b}.
We were involved in different IAEA CRP projects to produce recommended cross data for production $^{111g}$In, $^{114m}$In in \citep{Gul, Qaim, TF2001}. Production of recommended data for $^{110m}$In production is also included in the ongoing IAEA CRP \citep{Nichols}.
Searching of the literature for nuclear reaction data of proton induced reactions on indium showed only a few earlier experimental cross section and thick target yield data determinations. Cross section data were reported by \citep{Nortier1990} for production of $^{109,110,111,113}$Sn up to 100 MeV, by \citep{Nortier1991} for $^{109}$Sn, $^{109}$In, $^{109}$Cd up to 200 MeV, by \citep{Lundqvist} up to 85 MeV for production of $^{110}$Sn, by \citep{Musthafa} for $^{113}$Sn up to 20 MeV and by \citep{Johnson1979} for $^{115}$Sn up to 5.8 MeV.   Thick target yield data were reported by \citep{Dmitriev1981}  for $^{113}$Sn up to 22 MeV, by \citep{Dmitriev1982} for $^{113}$Sn and $^{114m}$In at 22 MeV and by \citep{Nickles} for $^{113m}$In and $^{115m}$In at 11 MeV.

\section{Experimental}
\label{2}

This study includes three irradiations at three different laboratories.  All cross section measurements of proton induced reactions on $^{nat}$In were done by using the activation method and stacked foil irradiation technique. Irradiations of commercial indium metal foil targets were performed at external beams of different cyclotrons.   
The first irradiation was done in 2006 at an external beam of CYRIC AVF 110 (Sendai, Japan) cyclotron at 70 MeV, for 1h at 58 nA beam intensity. The stack contained 13 stacked sets of Er(25 $\mu$m), Co(50 $\mu$m), Al(10 $\mu$m), In(50 $\mu$m) and Al(100 $\mu$m) foils. The covered energy range for Indium was 69.5-59.4 MeV. (The planned covered energy range was larger but by mistake the intended 500 $\mu$m Al absorbers were not inserted behind each set). 
The second irradiation was performed in 2014 at the LLN Cyclone 90 cyclotron (Belgium) with 65 MeV protons, for 1 h at 35 nA beam intensity. The stack contained 19 sets of 10 $\mu$m Al, 116 $\mu$m In, 99.2 $\mu$m Al, 8.41 $\mu$m V, 99.2 $\mu$m Al, 26.2 $\mu$m Ho and 99.2 $\mu$m Al. The Al foils in both irradiations served as monitors and to detect the possible recoils. The median proton energy in the last In-foil in the stack was 37.5 MeV.
The third stack was irradiated at the VUB CGR 560 (Brussels, Belgium) cyclotron at 35 MeV primary energy and 50 nA beam intensity for 1 h. The stack contained 16 sets of 10 $\mu$m Al, 116 $\mu$m In, 16 $\mu$m Au and 11 $\mu$m Ti foils. The Ti foils served as monitors. The covered energy range for the indium was 33.3-8.9 MeV. 

The stacks were mounted in a Faraday cup like target holder equipped with a long collimator. The activity of all produced radionuclides, both in the target material, and in the Al-backings/degraders and in the monitor foils were measured with HPGe detectors. The activity of the irradiated samples was measured nondestructively, without chemical separation. The source-detector distance was kept large enough to minimize dead time and pile up effects and possible influence of the non point-like geometry of the samples compared to calibration situations. Measurements were started 1 day after EOB in CYRIC due to the high radiation dose, at nearly EOB+10 h for irradiations at LLN due the high dose and to needed transport of the irradiated target to VUB. In case of these two high energy irradiations hence only longer-lived activation products could be identified. For the experiments at VUB also shorter-lived products could be identified as the first measurements started at EOB + 0.3 h.
Measurements were repeated a few times over a time period of several weeks to assess the low activities of longer-lived radioisotopes and to follow the decay of the produced radionuclides. In case of high energy, multi-target irradiations, the large number of different targets and the limited detector capacity made the optimization of gamma spectra measurements practically impossible. Gamma spectra were evaluated by using \citep{Canberra, Szekely} evaluation software. The complex spectra require iterative evaluation, by using theoretical predictions and preliminary experimental results.
The natural element indium contains two stable isotopes: $^{113}$In with 4.3 \% abundance and $^{115}$In with 95.7 \% abundance. Therefore, above some reaction thresholds only the so-called elemental cross-sections could be determined. Direct and cumulative cross-sections were calculated depending on the contributing processes and the activity measurements by using standard activation and decay formulas. The nuclear decay data used are collected in the Table 1 and were taken from \citep{Nudat}. For metastable states the energy level above ground is indicted in the first column. The reaction Q-values were determined by using the NNDC Q-value calculator \citep{Pritychenko}.

Number of incident particles was initially derived from the charge collected on the Faraday target holders in the three experiments. Correction of these values were made based on re-measurement of the cross sections of the monitor reactions $^{nat}$Al(p,x)$^{22,24}$Na over the whole energy range and comparison with recommended data (Fig. 1) taken from IAEA-TECDOC-1211 \citep{Gul}.
The incident beam energy on the targets was initially derived from accelerator settings and by calculation of the energy degradation in the target stack (Andersen and Ziegler, 1977). These results in the determination of incident energy were checked and corrected by assuring overlap of the excitation functions of the used monitor reactions with the recommended data according to \citep{TF2001}.
The number of the target nuclei was deduced from the thickness of the target. The thickness of individual targets was determined from the surface and of the mass of the target and the monitor foils.
The uncertainty on the cross section values was estimated by the standard technique according to the recommendations of the ISO guide (1993)\citep{Error}. The included experimental uncertainties are: number of target nuclei including non-uniformity (5 \%), incident particle flux (7 \%), peak area including statistical errors on counts (0.1-20 \%), detector efficiency (5 \%), $\gamma$-ray abundance and branching ratio data ($\>$1 \%, varies). Except for a few data points the total uncertainty of 10-15 \% was obtained as positive square root of the quadratic sum of the individual sources. Possible additional uncertainties due to non-linear effects of half-lives and waiting time were not accounted for. 
The monitoring method allows estimating the uncertainty on the primary beam energy at $\pm$ 0.2 MeV but the uncertainty on the average energy in each foil is continuously rising through the stack and attains $\pm$ 1.5 MeV for the last foil due to cumulative possible variations on target thickness and energy degradation.

\begin{table*}[t]
\tiny
\caption{Decay characteristic of the investigated reaction products}
\begin{center}
\begin{tabular}{|p{0.4in}|p{0.5in}|p{0.4in}|p{0.4in}|p{0.7in}|p{0.9in}|p{0.7in}|}
\hline
\textbf{Nuclide} & \textbf{Half-life} & \textbf{Decay method} &\textbf{E$_{\gamma}$(keV)} & \textbf{I$_{\gamma}$(\%)} & \textbf{Contributing 
reaction} & \textbf{Q-value(keV)} \\

\hline
$^{113m}$Sn\newline 7/2$^{+}$\newline77.382 keV & 21.4 min & EC 8.9 \%IT 91.1\% & 
77 & 0.501 & $^{113}$In(p,n)\newline $^{115}$\newline In(p,3n) & -1819.93\newline -18133.09 
\\
\hline
$^{113g}$Sn\newline 7/2$^{+}$ & 115.09 d & EC 100 \% & 255.134\newline 391.698 & 
2.11\newline 64.97 & $^{113}$In(p,n)\newline $^{115}$In(p,3n) & -1819.93\newline -18133.09 
\\
\hline
$^{111}$Sn\newline 7/2$^{+}$ & 35.3 min & EC 69.8 \%\newline  $\beta^{+}$ 30.25 \% & 
372.31\newline 761.97\newline 954.05\newline 1152.98\newline 1610.47 & 0.42\newline 1.48\newline 0.51\newline 2.7\newline 1.31 & $^{113}$
In(p,3n)\newline $^{115}$In(p,5n) & -20351.3\newline -36664.46 \\
\hline
$^{110}$Sn\newline 0$^{+}$ & 4.11 h & EC 100 \% & 280.462 & 100 & $^{113
}$In(p,4n)\newline $^{115}$In(p,6n) & -28520.1\newline -44833.3 \\
\hline
$^{109}$Sn\newline 5/2$^{+}$ & 18 min & EC 93.6 \%\newline  $\beta^{+}$ 6.6 \% & 
649.8\newline 1099.2\newline 1321.3 & 28\newline 30\newline 11.9 & $^{113}$In(p,5n)\newline $^{115}$In(p,7n) & 
-39802.54\newline -56115.7 \\
\hline
$^{115m}$In\newline 1/2$^{-}$\newline 336.244 keV & 4.486 h & IT: 95.0 \% $\beta^{-}$
: 5.0 \% & 336.24 & 45.8 & $^{115}$In(p,p) & 0.0 \\
\hline
$^{114m}$In\newline 5$^{+}$\newline 190.368 keV & 49.51 d & EC 3.25 \%\newline IT 96.75 \% & 
190.27\newline 558.43\newline 725.24 & 15.56\newline 3.2  3.2  & $^{115}$In(p,pn) & 
-9039.26 \\
\hline
$^{113m}$In\newline 1/2$^{-}$\newline 391.691 keV & 99.476 min & IT 100 \% & 
391.698 & 64.94 \% & $^{113}$In(p,p)\newline $^{115}$In(p,p2n) & - 
0.0\newline -16313.16 \\
\hline
$^{111}$In\newline 9/2$^{+}$ & 2.8047 d & $\varepsilon$: 100 & 171.28\newline 245.35 & 90.7 
\newline 94.1  & $^{113}$In(p,p2n)\newline $^{115}$In(p,p4n)\newline $^{111}$Sn 
decay & -17117.55\newline -33430.73\newline -20351.3 \\
\hline
$^{110m}$In\newline 2$^{+}$\newline 62.08 keV & 69.1 min & $\beta^{+}$: 61.3\%\newline  EC 
:38.7\% & 657.75 & 97.74 & $^{113}$In(p,p3n)\newline $^{115}$In(p,p5n)\newline $^{
110}$Sn decay & -27109.0\newline -43422.2\newline -28520.1 \\
\hline
$^{110g}$In\newline 7$^{+}$ & 4.92 h & $\beta^{+}$ :0.0081 \%\newline EC:.99.9919 \% 
& 641.68\newline 657.75\newline 707.40\newline 937.478\newline 997.16 & 26\newline 98\newline 29.5\newline 68.4\newline 10.5 & $^{113}$
In(p,p3n)\newline $^{115}$In(p,p5n) & -27109.0\newline -43422.2 \\
\hline
$^{109}$In & 4.167 h & EC:95.44\%\newline $\beta^{+:}$ 4.56\% & 203.5\newline 426.2\newline 623.5 
& 73.5\newline 4.12\newline 5.5 & $^{113}$In(p,p4n)\newline $^{115}$In(p,p6n) & 
-35163.17\newline -51476.34 \\
\hline
$^{111m}$Cd\newline 11/2$^{-}$\newline 396.22 keV & 48.54 min & IT & 150.825\newline 245.395 
& 29.1\newline  94 & $^{113}$In(p,2pn)\newline $^{115}$In(p,2p3n)\newline $^{111}$In 
decay & -15473.03\newline -31786.195\newline -17117.55 \\
\hline
$^{109}$Cd\newline 5/2$^{+}$ & 461.4 d & $\varepsilon$: 100 \% & 88.0336 & 3.7 & $^{
113}$In(p,2p3n)\newline $^{115}$In(p,2p5n)\newline $^{109}$In decay & 
-32364.39\newline -48677.55\newline -35163.17 \\
\hline
\end{tabular}

\end{center}
\begin{flushleft}
\tiny{\noindent The Q-values shown in Table 1 refer to the formation of the ground state. Increase absolute Q-values for isomeric states with level energy of the isomer. When complex particles are emitted instead of individual protons and neutrons the absolute Q-values have to be decreased by the respective binding energies of the compound particles: np-d, +2.2 MeV; 2np-t, +8.48 MeV; 2p2n-$\alpha$, +28.30 MeV.
}
\end{flushleft}

\end{table*}




\begin{figure}
\includegraphics[width=0.5\textwidth]{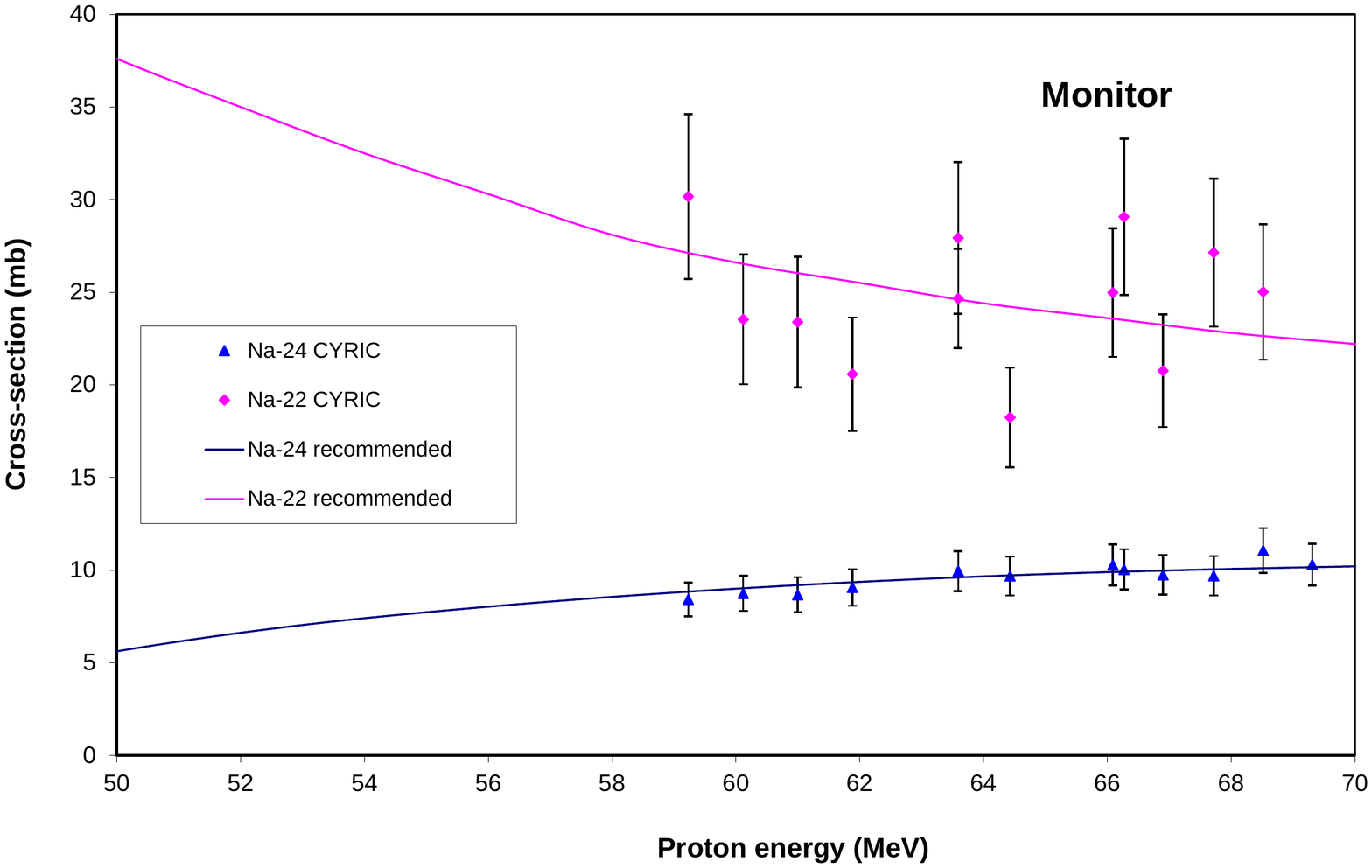}
\caption{Excitation functions of the $^{27}$Al(p,x)$^{22,24}$Na monitor reactions}
\label{fig:1}       
\end{figure}

\section{Results}
\label{3}

\subsection{Production cross sections}
\label{3.1}
The new experimental data are discussed for the different activation products separately in the next paragraphs. Numerical values for the cross sections (direct or cumulative) are presented in Tables 2-3 and in Figures 2-10. Results of different irradiations will be presented separately in the figures. The new experimental results are compared with the data found in the literature and with the theoretical calculations in the TENDL-2014 \citep{Koning2013} TALYS based \citep{Koning2012} library. The cross sections were evaluated using the average of the strong independent gamma-lines (when those were consistent) for every investigated radionuclide. The selection criteria of gamma-lines involved in the average calculation were: no interference with other gamma-lines/isotopes; suitable statistic, which means that not always the same set of gamma-lines could be used for the same isotope in the different experiments.

\subsubsection{Production of $^{113}$Sn}
\label{3.1.1}
The $^{113}$Sn isotope has a long-lived (115.09 d) ground state and a short-lived metastable state (21.4 min), which could not be assessed reliably in our experiments. We present hence the cumulative production of $^{113g}$Sn after total decay of the isomeric state (IT 91.1 \%). The $^{113}$In(p,n) and $^{115}$In(p,3n) reactions can contribute to the formation of this isotope and we can hence observe on Fig. 2 an excitation function with two expected  maxima. According to Fig. 2 there is a good agreement between all experimental data and the theoretical results. 

\begin{figure}
\includegraphics[width=0.5\textwidth]{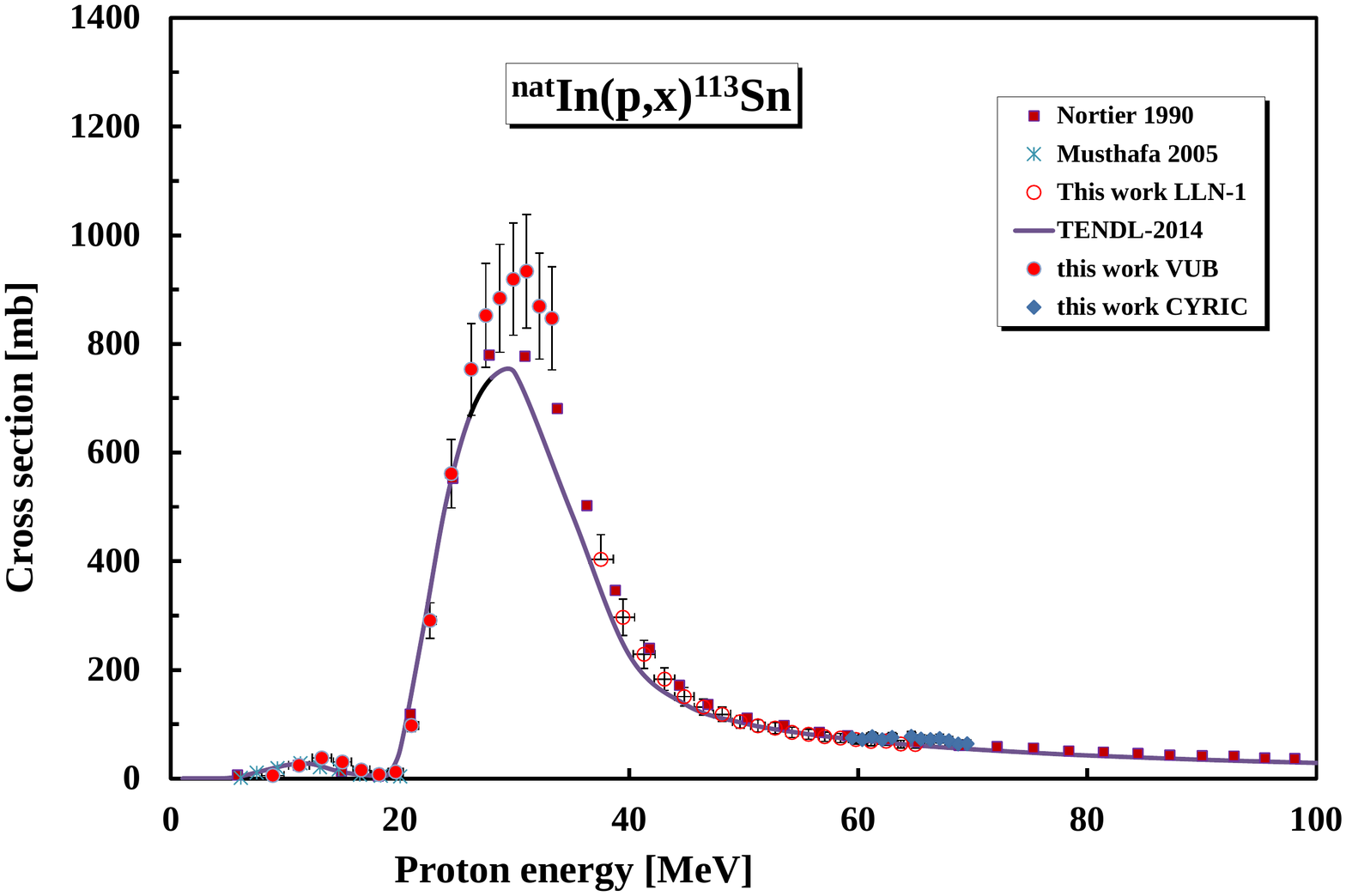}
\caption{Excitation functions of the $^{nat}$In(p,xn)$^{113}$Sn reaction}
\label{fig:2}       
\end{figure}

\subsubsection{Production of $^{111}$Sn}
\label{3.1.2}
New cross section data for production of the $^{111}$Sn (35.3 min, 7/2$^+$) were obtained only from the low energy irradiation (VUB), where the reduced dose and the logistic circumstances allowed to use short cooling time after EOB. The cross sections of $^{113}$In(p,3n) reaction were hence measured practically up to the threshold of the $^{115}$In(p,5n) reaction (Q = -36664.46 keV). The agreement with the earlier experimental data and with the theory is good (see Fig. 3).

\begin{figure}
\includegraphics[width=0.5\textwidth]{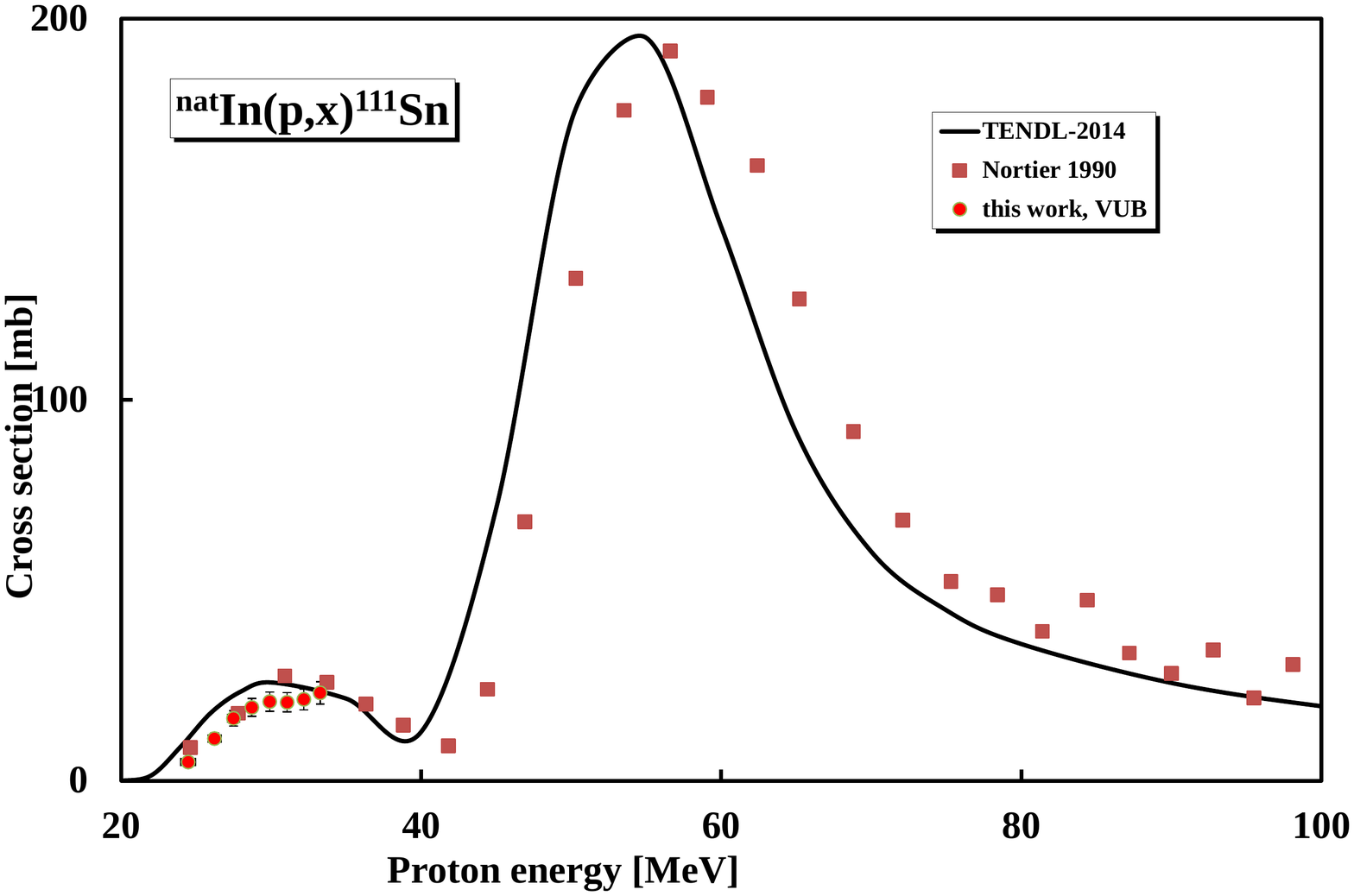}
\caption{Excitation functions of the $^{nat}$In(p,xn)$^{111}$Sn reaction in comparison with the model  calculations}
\label{fig:3}       
\end{figure}

\subsubsection{Production of $^{110}$Sn}
\label{3.1.3}
The excitation functions for production of $^{110}$Sn (4.11 h) are shown in Fig. 4 in comparison with the earlier experimental data of \citep{Nortier1990} and \citep{Lundqvist}. The agreement in the whole energy range is good, except the last 4 point of the CYRIC results, which was possibly caused by the inhomogeneity in the target foils accumulating a systematic shift towards the end of the stack. The theory underestimates the cross section of the $^{115}$In(p,6n) reaction at high energies.

\begin{figure}
\includegraphics[width=0.5\textwidth]{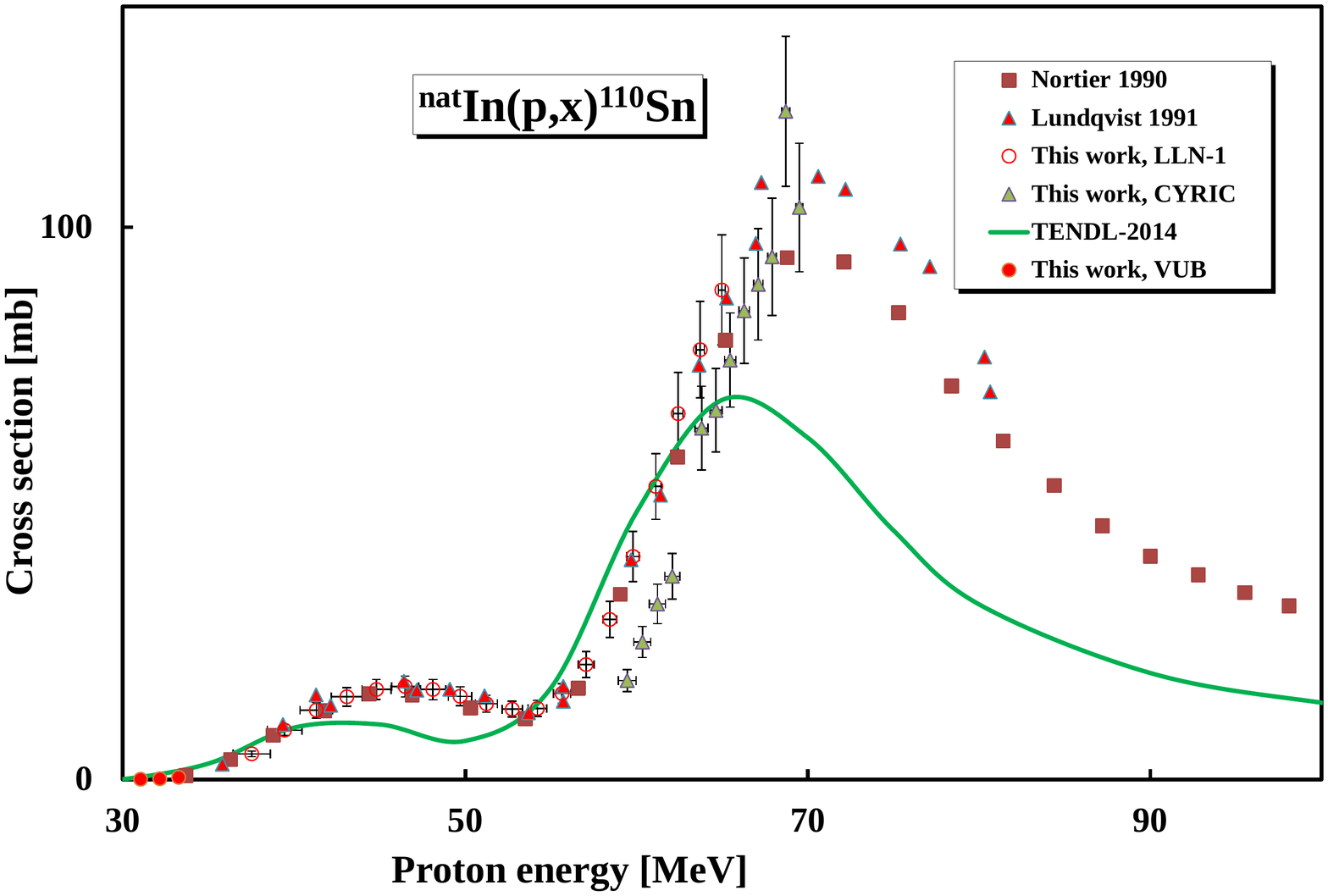}
\caption{Excitation functions of the $^{nat}$In(p,xn)$^{110}$Sn reaction}
\label{fig:4}       
\end{figure}
\vspace{1 cm}

\subsubsection{Production of $^{115m}$In}
\label{3.1.4}
The $^{115}$In isotope has a very long-lived (4.41E+14 a) ground state and one shorter-lived metastable state (4.486 h). We present cross section data for the production of the isomeric state (Fig. 5) decaying overwhelmingly (IT = 95 \%) by internal transition to the ground state. In the investigated energy range the $^{115}$In(p,p$^,$) reaction is  the only contributing reaction. Our experimental results follow the TENDL prediction in shape but the magnitude is higher. The decay of the long-lived ground state could not be assessed in our experiments.

\begin{figure}
\includegraphics[width=0.5\textwidth]{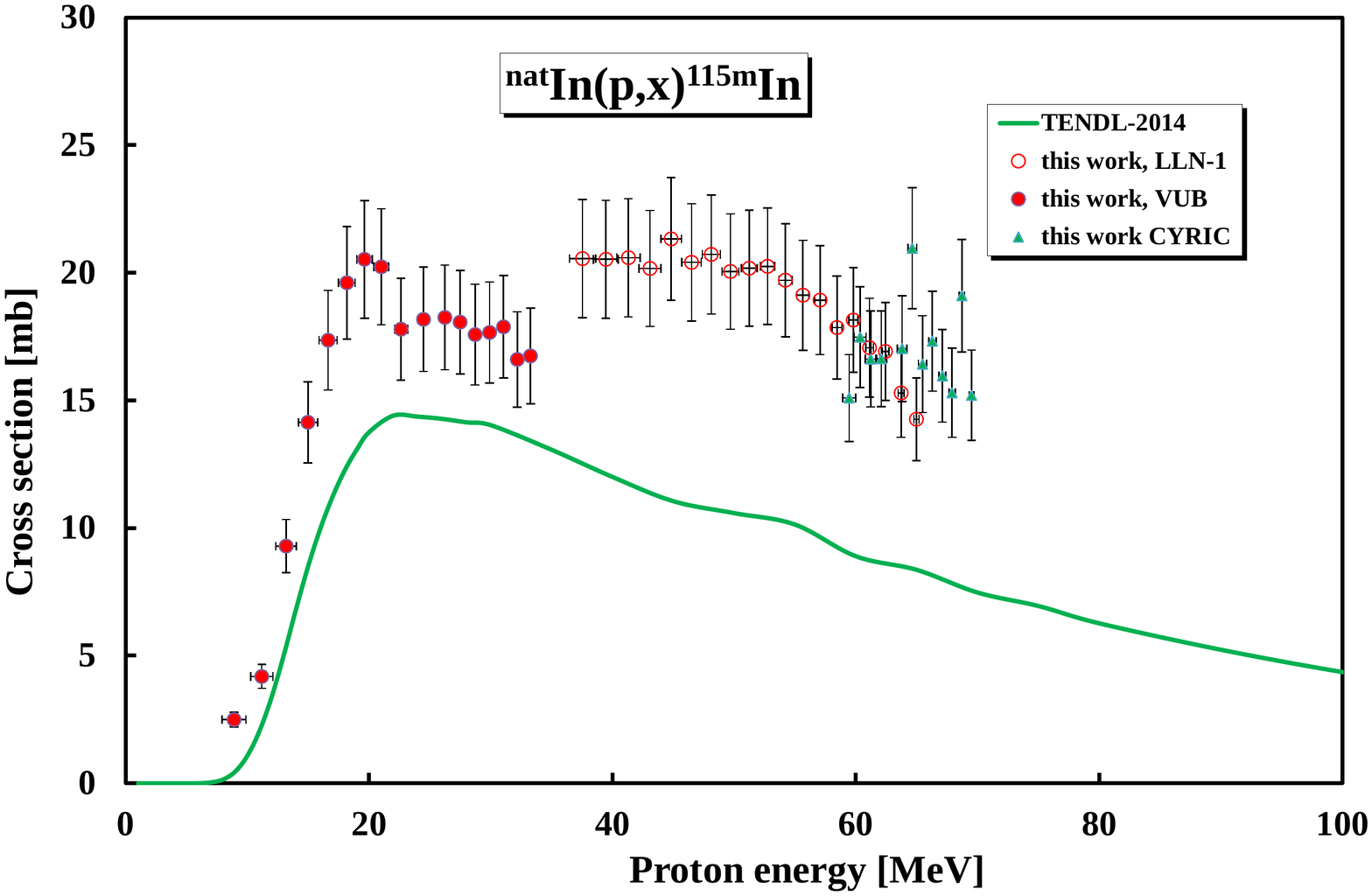}
\caption{Excitation functions of the $^{nat}$In(p,x)$^{115m}$In reaction in comparison with the model calculations}
\label{fig:5}       
\end{figure}

\subsubsection{Production of $^{114m}$In}
\label{3.1.5}
The radioisotope $^{114}$In has a short-lived ground state (71.9 s) and a long-lived (49.51 d) metastable state, which decays with 96.7 \% by IT to the ground state and with 3.3 \% by EC to stable $^{114}$Cd. Only the production cross section of the longer-lived metastable state was determined. According to Fig. 6 the measured and theoretical data are similar in shape, but TENDL-2014 overestimates the experimental cross section data at higher energies. The production of $^{114m}$In is the result of direct (p,pn) reaction on $^{115}$In.

\begin{figure}
\includegraphics[width=0.5\textwidth]{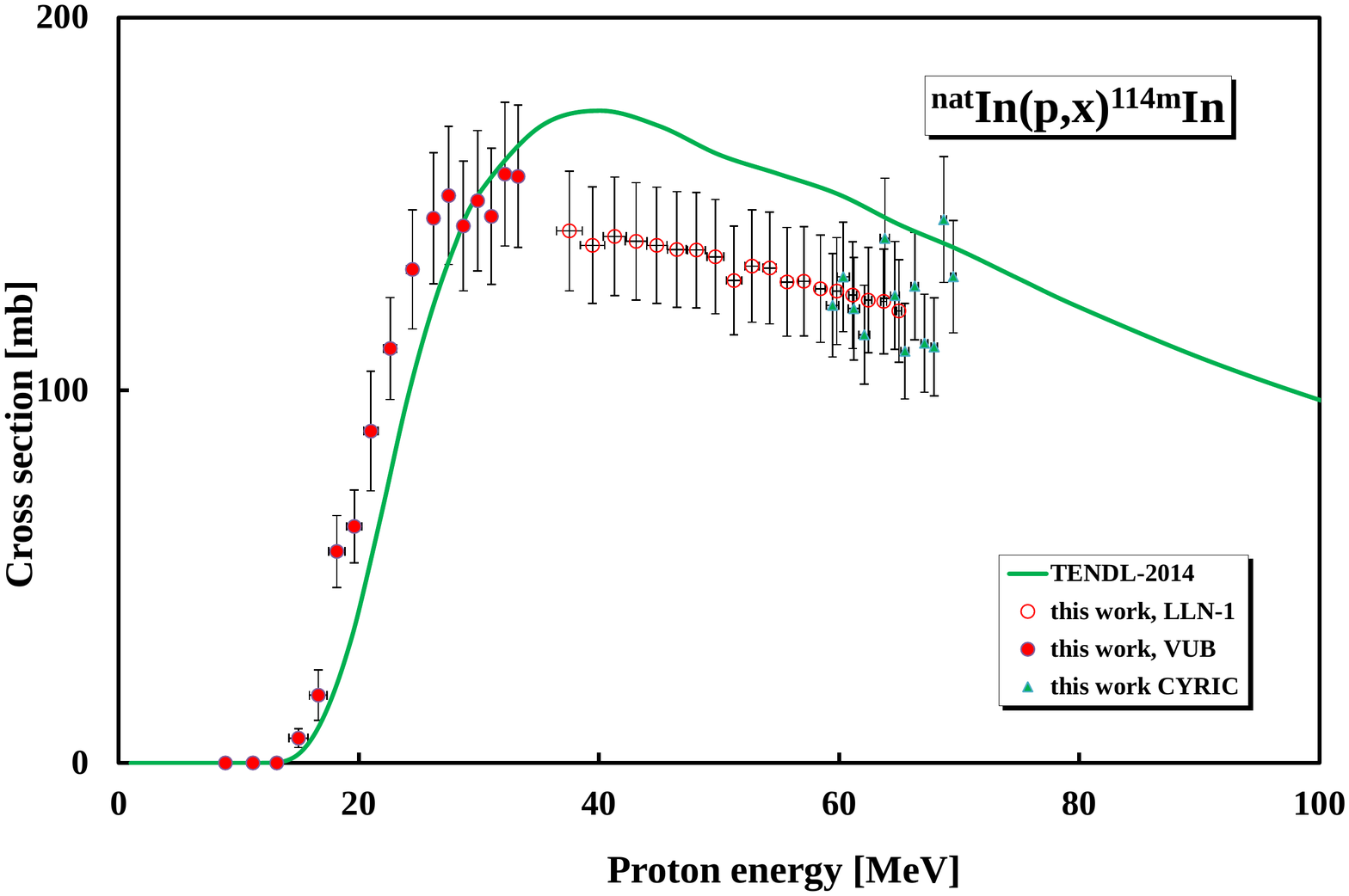}
\caption{Excitation functions of the $^{nat}$In(p,x)$^{114m}$In reaction}
\label{fig:6}       
\end{figure}

\subsubsection{Production of $^{113m}$In}
\label{3.1.6}
The 99.476 min half-life metastable state of $^{113}$In can be produced only directly via the $^{115}$In(p,p2n) reaction and from the decay of $^{113g}$Sn. The metastable state decays by 100 \% IT to the stable ground state 113gIn. For measurements shortly after EOB in growth from the long-lived parent $^{113}$Sn (115 d) can be neglected (the $^{113m}$Sn EC decay leads to $^{113}$In ground state) and hence we present here the direct formation. In average there is a good agreement in the shape between the experimental data and the theoretical calculations. The TENDL-2014 slightly, but systematically overestimates the experiment (Fig. 7).

\begin{figure}
\includegraphics[width=0.5\textwidth]{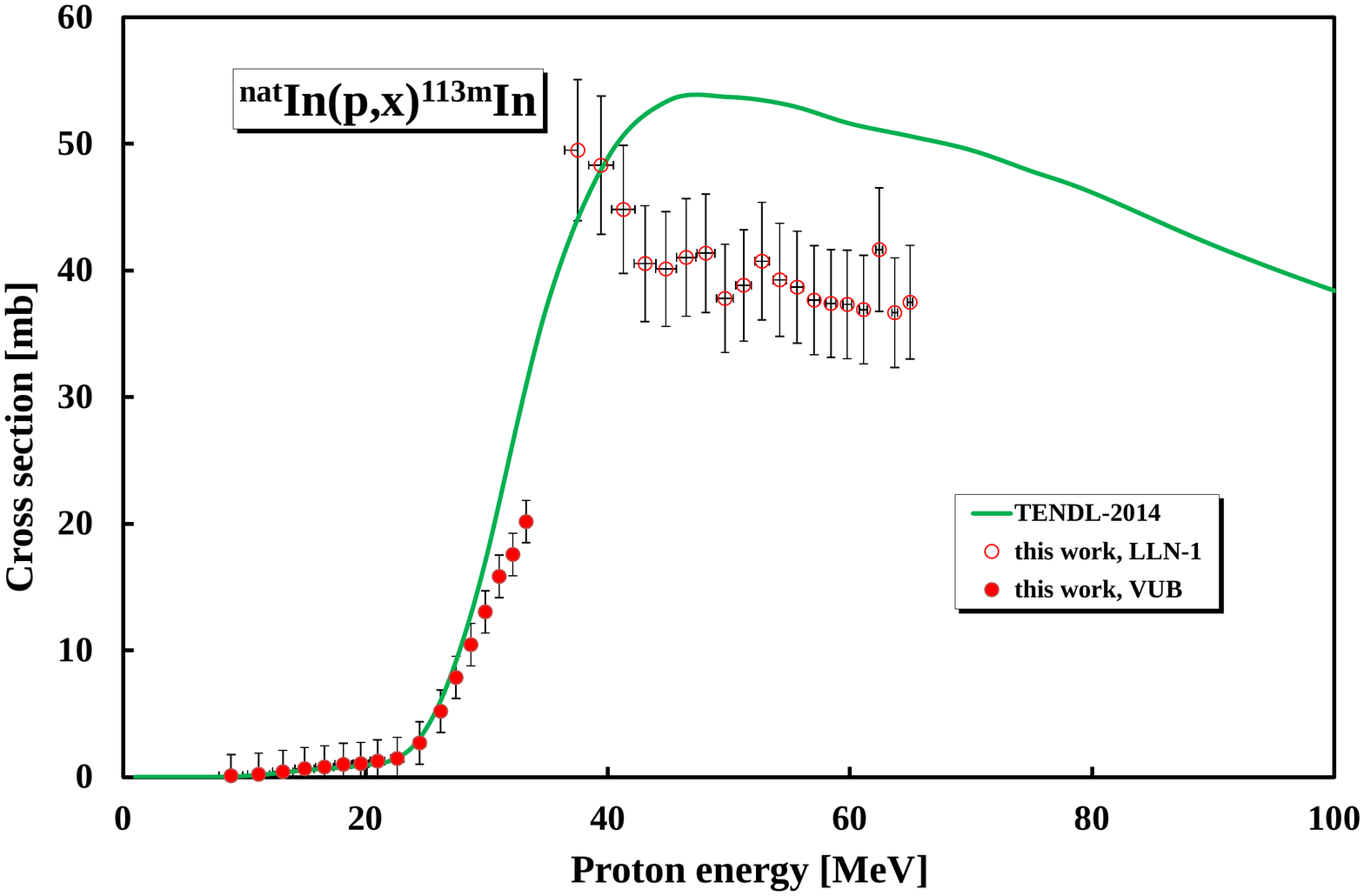}
\caption{Excitation functions of the $^{nat}$In(p,x)$^{113m}$In reaction in comparison with the model calculations}
\label{fig:7}       
\end{figure}

\subsubsection{Production of $^{112m}$In}
\label{3.1.7}
Out of the two isomeric states we could deduce production cross sections for the longer-lived, higher lying state (20.67 min, 156.613 keV) decaying by IT 100 \% to the 14.88 min ground state (Fig. 8) in the low energy irradiation. There is good correspondence between the experiment and the theory.

\begin{figure}
\includegraphics[width=0.5\textwidth]{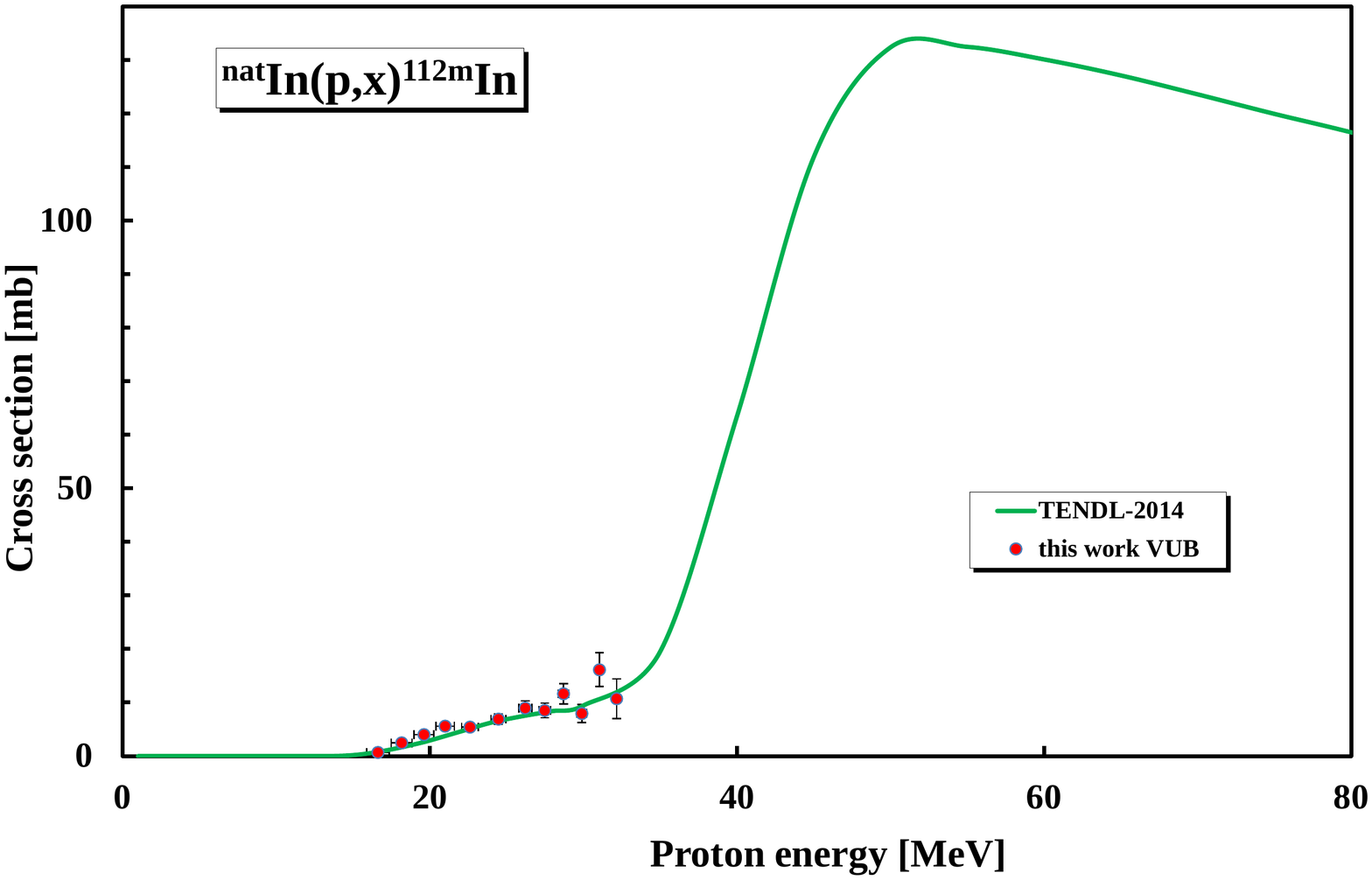}
\caption{Excitation functions of the $^{nat}$In(p,x)$^{112m}$In reaction in comparison with the model calculations}
\label{fig:8}       
\end{figure}

\subsubsection{Production of $^{111g}$In}
\label{3.1.8}
The $^{111}$In radionuclide has a short-lived (T$_{1/2}$= 7.7 min) metastable state decaying to the 2.83 d ground state by IT. The cumulative production after total decay of the metastable state was measured. The formation of the ground state of $^{111}$In contains three contributions: direct production by (p,pxn) reactions, decay of the short-lived metastable state (isomeric transition ) and decay of the short-lived parent $^{111}$Sn (35.3 min). In average the agreement between the experimental data and the theoretical calculations is satisfactory (Fig. 9).

\begin{figure}
\includegraphics[width=0.5\textwidth]{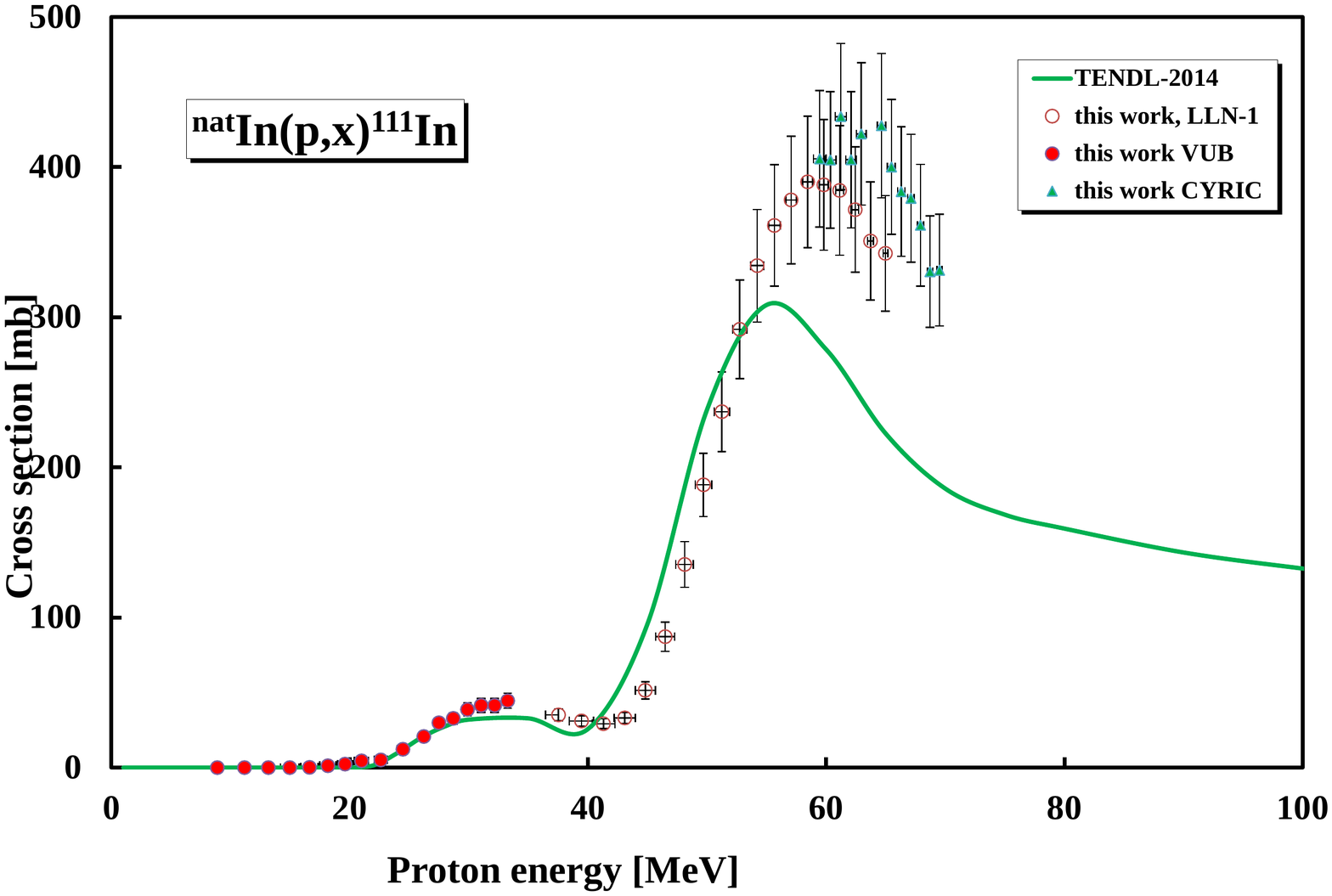}
\caption{9.	Excitation functions of the $^{nat}$In(p,x)$^{111g}$In cumulative reaction}
\label{fig:9}       
\end{figure}

\subsubsection{Production of $^{110g}$In and $^{110m}$In}
\label{3.1.9}
The shorter-lived isomeric state (T$_{1/2}$= 69.1 min, 62.08 keV) decays not into the longer-lived ground state (T$_{1/2}$= 4.92 h). The parent $^{110}$Sn decays only to the isomeric state. The cross sections for short-lived isomeric state could not be deduced in our high energy experiments due to the long cooling time. The shorter cooling times for the VUB experiment should in principle have allowed to deduce cross sections for production of both states, taking into account the contribution from the $^{110}$Sn decay in case of $^{110m}$In. The gamma-lines of the ground and isomeric states are however practically identical. We hence only determined the cumulative production cross section for the ground state from spectra taken after longer cooling times (Fig. 10). The agreement is acceptable, except the last 4 point of the CYRIC results, which was possibly caused by the inhomogeneity in the target foils accumulating a systematic shift towards the end of the stack.

\begin{figure}
\includegraphics[width=0.5\textwidth]{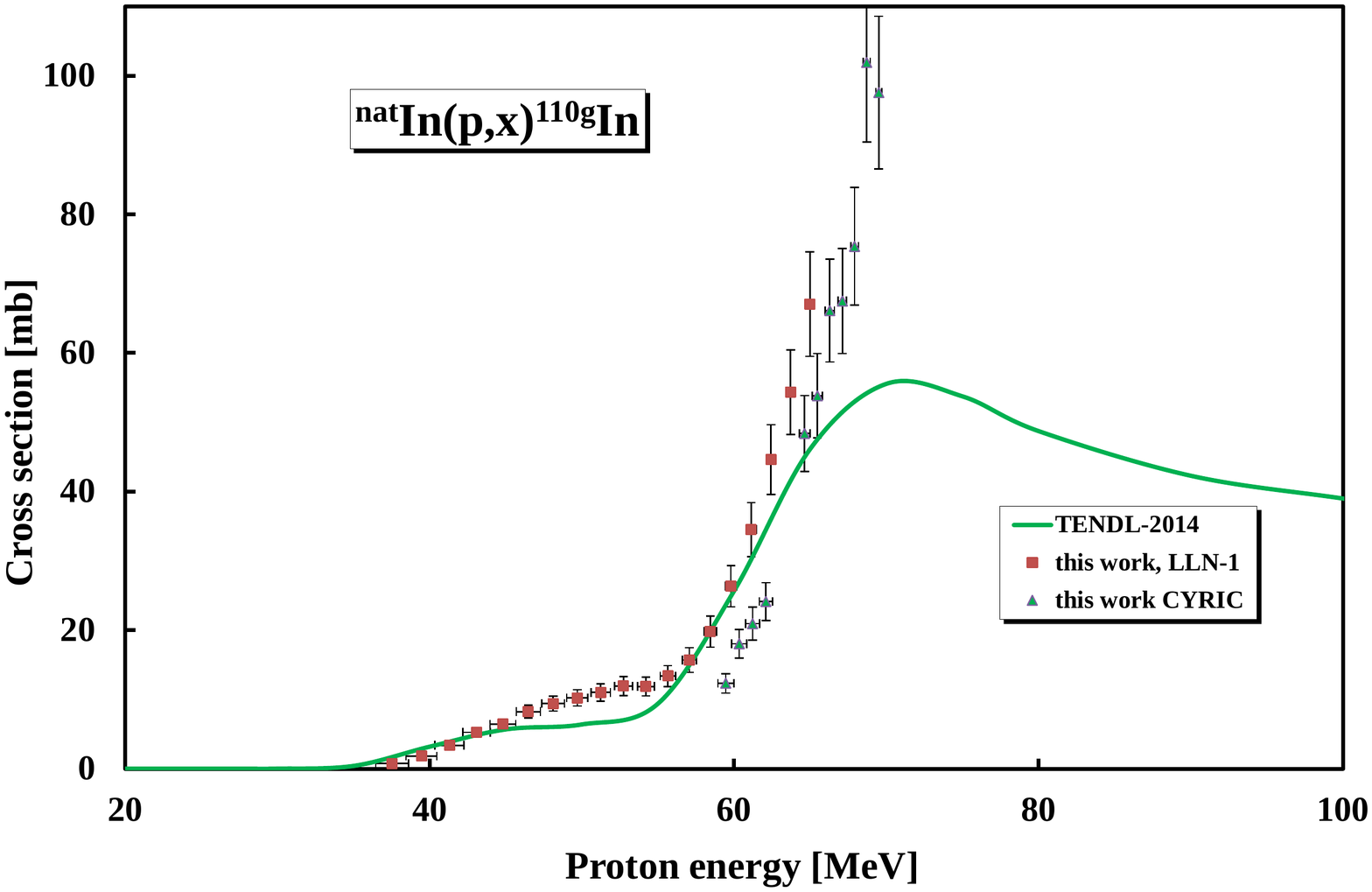}
\caption{Excitation functions of the $^{nat}$In(p,x)$^{110g}$In reaction}
\label{fig:10}       
\end{figure}

\subsubsection{Production of $^{109}$In}
\label{3.1.10}
The measured cumulative cross sections for production of $^{109g}$In (4.167 h ) contains the contribution from total decay of the $^{109m}$In isomeric state (1.34 min, IT: 100 \%) and  from the decay of the $^{109}$Sn parent isotope (18.0 min, $\varepsilon$: 100 \%) (Fig. 11). Rather good agreement was found between the values of this experiment and \citep{Nortier1990} and the TENDL-2014 library.

\begin{figure}
\includegraphics[width=0.5\textwidth]{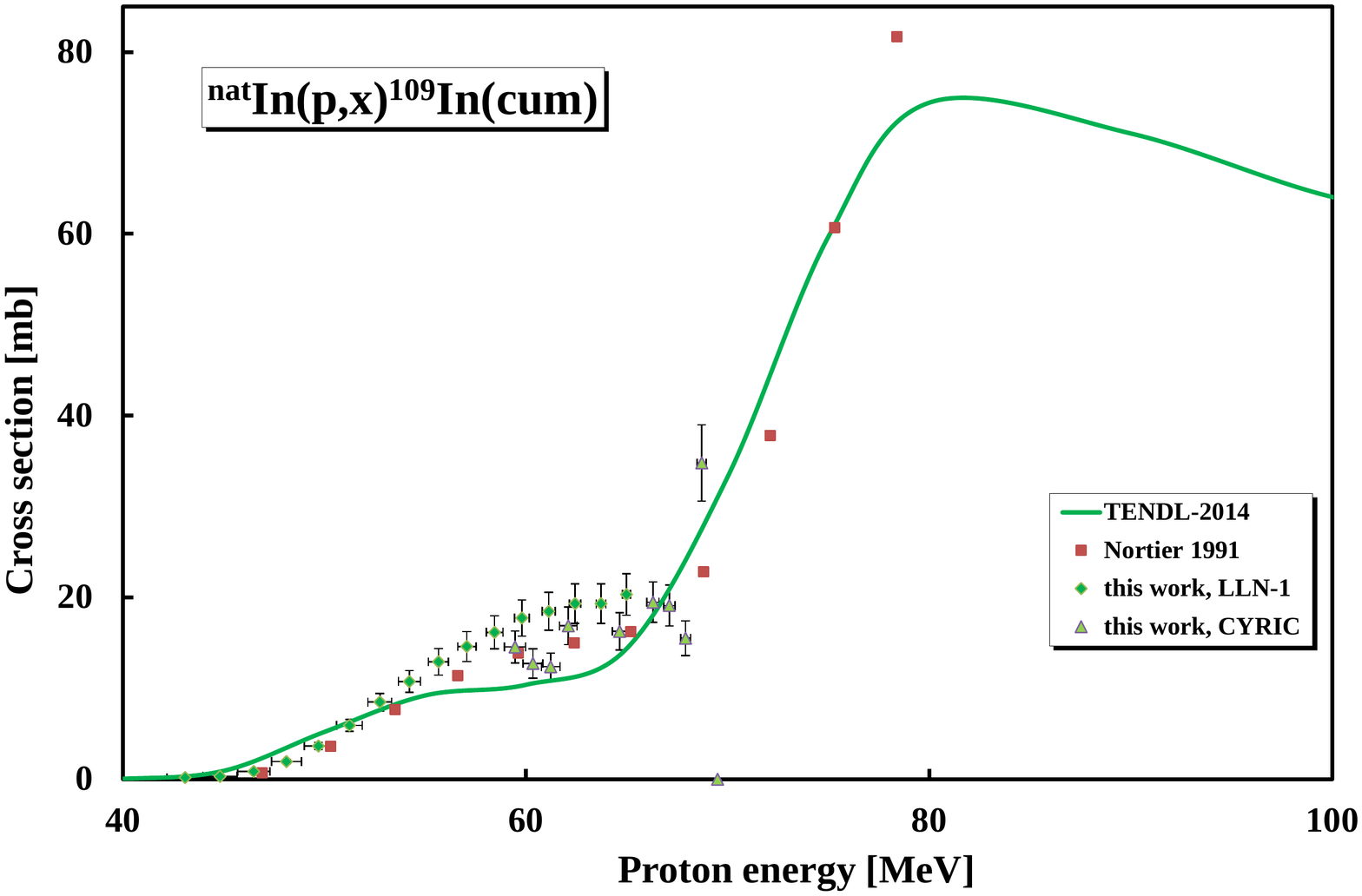}
\caption{Excitation functions of the $^{nat}$In(p,x)$^{109}$In reaction}
\label{fig:11}       
\end{figure}

\subsubsection{Production of $^{111m}$Cd}
\label{3.1.11}
The $^{111}$Cd has a stable ground state and a 48.54 min metastable state. In principle two routes are possible for production of the metastable state: direct production by In(p,2pxn) reaction and decay chain of the $^{111}$Sn (-22574.38 keV) -$^{111}$In (100 \%)-$^{111m}$Cd(0.005 \%), which is negligible. As our presented data were obtained from spectra measured a few hours after EOB (short half-life of $^{111m}$Cd), the ingrowth from the decay of the longer-lived $^{111g}$In is in any case negligible and the cross sections reflect the direct production. The measured and calculated cross sections are shown in Fig. 12.

\begin{figure}
\includegraphics[width=0.5\textwidth]{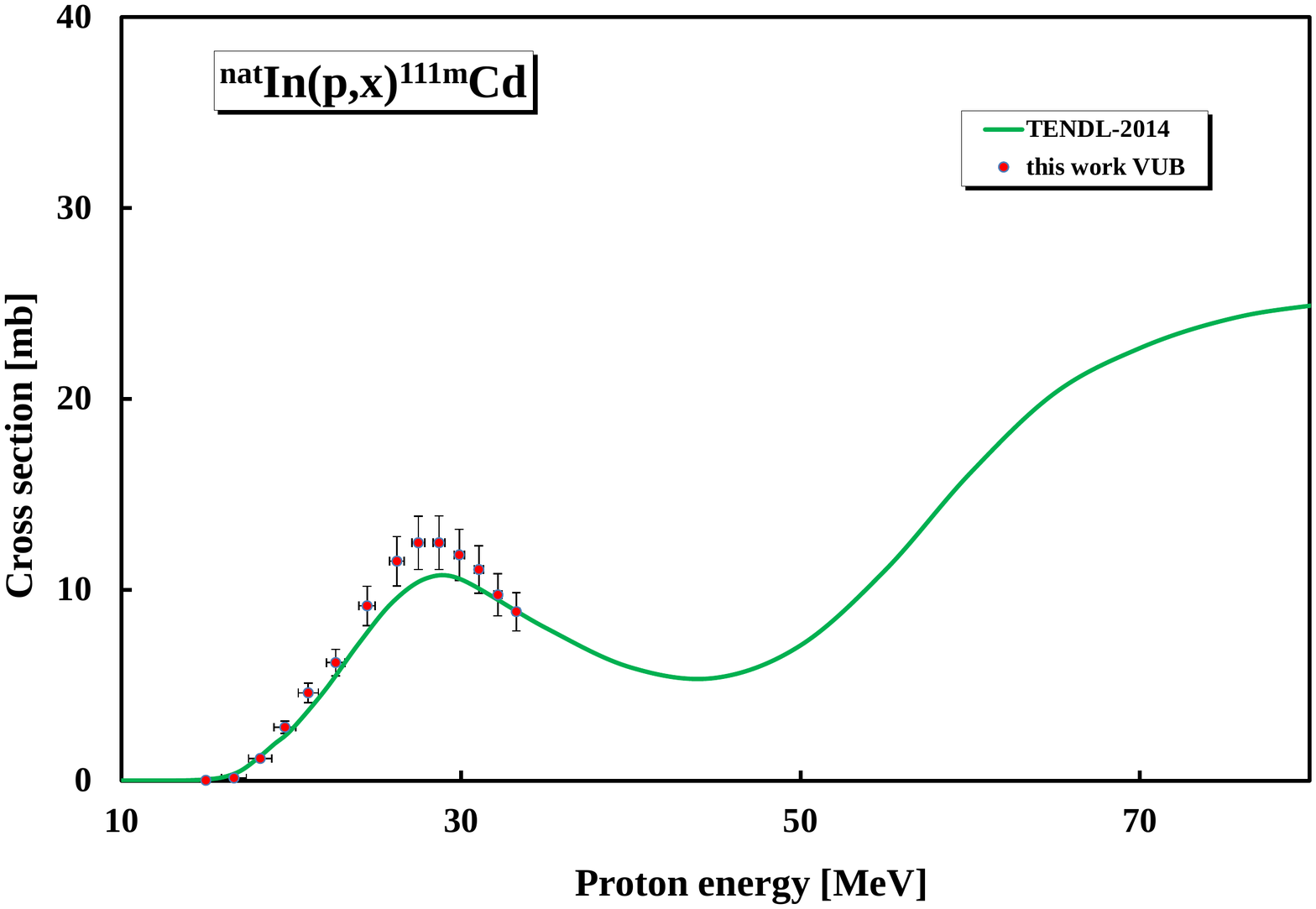}
\caption{Excitation functions of the $^{nat}$In(p,x)$^{111m}$Cd reaction in comparison with the model calculations}
\label{fig:12}       
\end{figure}

\subsubsection{Production of $^{109}$Cd}
\label{3.1.12}
In Fig. 13 the cumulative cross sections of long-lived $^{109}$Cd (461.4 d) are shown, including direct production and possible contributions through the $^{109}$Sn (18.0 min, ε: 100 \%)-$^{109}$In (4.167 h, EC 95.44 \% and $\beta^+$ 4.56 \%)-$^{109}$Cd decay chain.

\begin{figure}
\includegraphics[width=0.5\textwidth]{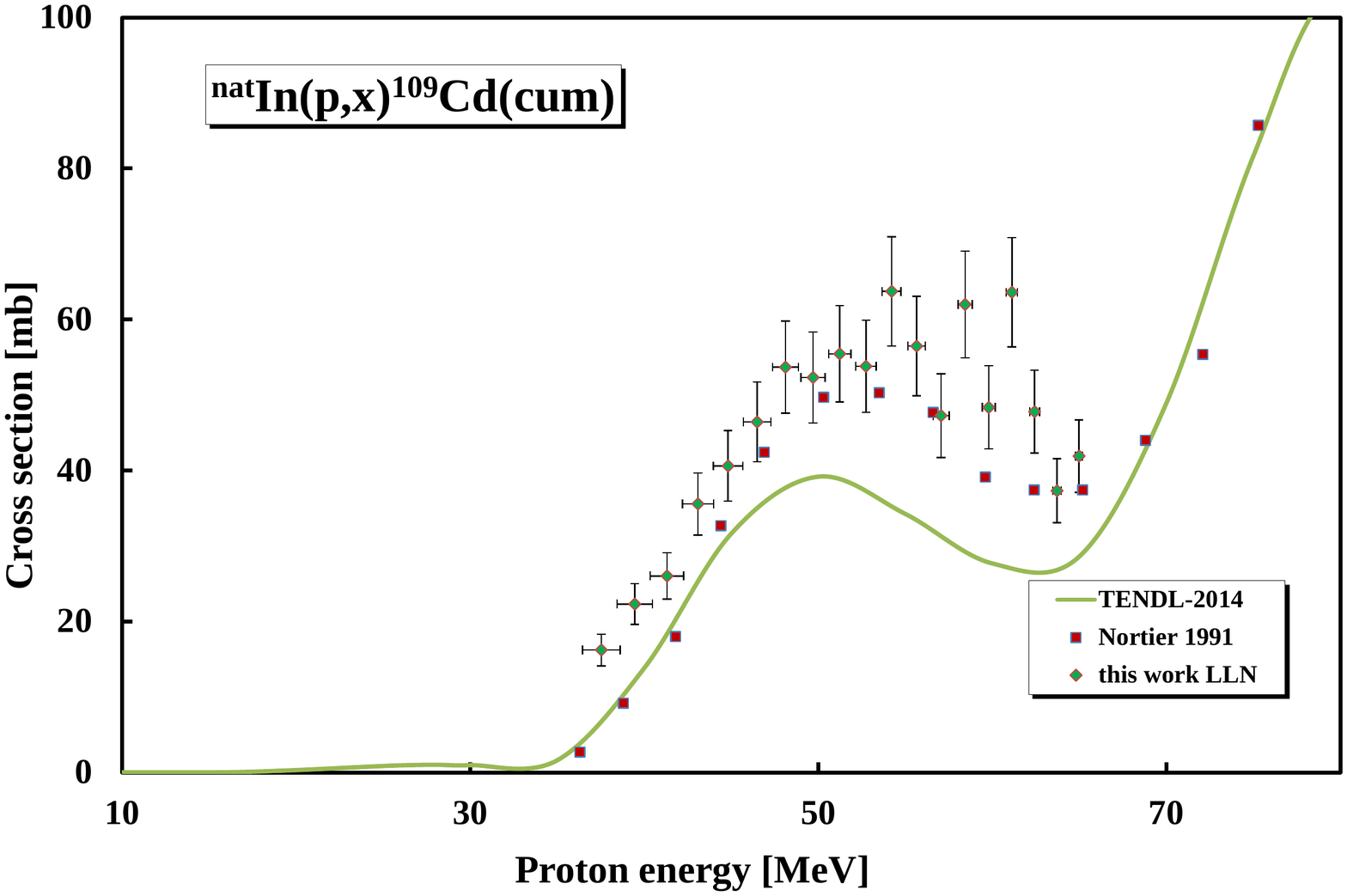}
\caption{Excitation functions of the $^{nat}$In(p,x)$^{109}$Cd reaction in comparison with the model calculations}
\label{fig:13}       
\end{figure}

\begin{table*}[t]
\tiny
\caption{Measured cross sections of the $^{nat}$In(d,x)$^{113g,111,110}$Sn, $^{115m,114m,113m}$In reactions and the estimated mean uncertainties}
\begin{center}
\begin{tabular}{|l|l|l|l|l|l|l|l|l|l|l|l|l|l|l|}
\hline
 & \textbf{Energy} & \textbf{$\Delta$E} & \multicolumn{2}{|c|}{\textbf{$^{113g}$Sn}} & 
\multicolumn{2}{|c|}{\textbf{$^{111}$Sn}} & \multicolumn{2}{|c|}{\textbf{$^{110}$Sn}} & \multicolumn{2}{|c|}{\textbf{$^{115m
}$In}} & \multicolumn{2}{|c|}{\textbf{$^{114m}$In}} & \multicolumn{2}{|c|}{\textbf{$^{113m}$In}} \\
\cline{2-15}
 &\multicolumn{2}{|c|}{} & \textbf{$\sigma$} & \textbf{$\pm\Delta\sigma$} & \textbf{$\sigma$} & \textbf{$\pm\Delta\sigma$} & 
\textbf{$\sigma$} & \textbf{$\pm\Delta\sigma$} & \textbf{$\sigma$} & \textbf{$\pm\Delta\sigma$} & \textbf{$\sigma$
} & \textbf{$\pm\Delta\sigma$} & \textbf{$\sigma$} & \textbf{$\pm\Delta\sigma$} \\
\cline{4-15}
 & \multicolumn{2}{|c|}{\textbf{MeV}} & \multicolumn{12}{|c|}{\textbf{mbarn}} \\
\hline
\textbf{LLN} & 37.5 & 1.1 & 403.5 & 45.3 & & & 4.6 & 0.5 & 20.6 & 2.3 
& 142.7 & 16.1 & 49.5 & 5.6 \\
\cline{2-15}
 & 39.5 & 1.0 & 296.6 & 33.3 & & & 8.9 & 1.0 & 20.5 & 2.3 & 138.9 & 15.6 
& 48.3 & 5.5 \\
\cline{2-15}
 & 41.3 & 1.0 & 228.5 & 25.7 & & & 12.5 & 1.4 & 20.6 & 2.3 & 141.3 & 15.9 
& 44.8 & 5.1 \\
\cline{2-15}
 & 43.1 & 0.9 & 182.9 & 20.6 & & & 14.9 & 1.7 & 20.2 & 2.3 & 140.0 & 15.7 
& 40.6 & 4.6 \\
\cline{2-15}
 & 44.8 & 0.8 & 150.7 & 16.9 & & & 16.3 & 1.8 & 21.3 & 2.4 & 138.9 & 15.6 
& 40.1 & 4.5  \\
\cline{2-15}
 & 46.5 & 0.8 & 131.5 & 14.8 & & & 16.8 & 1.9 & 20.4 & 2.3 & 137.8 & 15.5 
& 41.0 & 4.6 \\
\cline{2-15}
 & 48.1 & 0.7 & 118.1 & 13.3 & & & 16.3 & 1.8 & 20.7 & 2.3 & 137.6 & 15.5 
& 41.4 & 4.7 \\
\cline{2-15}
 & 49.7 & 0.7 & 104.4 & 11.7 & & & 15.1 & 1.7 & 20.0 & 2.3 & 135.8 & 15.3 
& 37.8 & 4.3 \\
\cline{2-15}
 & 51.2 & 0.6 & 96.9 & 10.9 & & & 13.7 & 1.5 & 20.2 & 2.3 & 129.4 & 14.6 
& 38.8 & 4.4 \\
\cline{2-15}
 & 52.7 & 0.6 & 92.5 & 10.4 & & & 12.7 & 1.4 & 20.3 & 2.3 & 133.3 & 15.1 
& 40.7 & 4.6 \\
\cline{2-15}
 & 54.2 & 0.5 & 84.9 & 9.6 & & & 12.8 & 1.4 & 19.7 & 2.2 & 132.8 & 15.0 & 
39.3 & 4.5 \\
\cline{2-15}
 & 55.7 & 0.5 & 81.3 & 9.2 & & & 15.6 & 1.8 & 19.1 & 2.2 & 129.1 & 14.6 & 
38.7 & 4.4 \\
\cline{2-15}
 & 57.1 & 0.5 & 76.9 & 8.7 & & & 20.8 & 2.3 & 18.9 & 2.1 & 129.2 & 14.6 & 
37.7 & 4.3 \\
\cline{2-15}
 & 58.4 & 0.4 & 74.2 & 8.4 & & & 29.0 & 3.3 & 17.9 & 2.0 & 127.2 & 14.4 & 
37.4 & 4.3 \\
\cline{2-15}
 & 59.8 & 0.4 & 71.7 & 8.1 & & & 40.4 & 4.5 & 18.2 & 2.0 & 126.6 & 14.3 & 
37.3 & 4.3 \\
\cline{2-15}
 & 61.1 & 0.3 & 68.2 & 7.7 & & & 53.0 & 6.0 & 17.1 & 1.9 & 125.5 & 14.3 & 
36.9 & 4.3 \\
\cline{2-15}
 & 62.4 & 0.3 & 68.5 & 7.7 & & & 66.3 & 7.4 & 16.9 & 1.9 & 124.2 & 14.1 & 
41.6 & 4.9 \\
\cline{2-15}
 & 63.7 & 0.2 & 63.3 & 7.1 & & & 77.8 & 8.7 & 15.3 & 1.7 & 123.8 & 14.0 & 
36.7 & 4.3 \\
\cline{2-15}
 & 65.0 & 0.2 & 62.5 & 7.1 & & & 88.6 & 10.0 & 14.3 & 1.6 & 121.3 & 13.8 
& 37.5 & 4.5 \\
\cline{2-15}
 & & &  & &  & &  & &  & &  & &  & \\
\hline
\textbf{VUB} & 8.9 & 1.0 & 5.5 & 0.6 & & & & & 2.5 & 0.3 & & & 0.1 & 
0.01 \\
\cline{2-15}
 & 11.2 & 0.9 & 24.4 & 2.8 & & & & & 4.2 & 0.5 & & & 0.2 & 0.03 \\
\cline{2-15}
 & 13.2 & 0.8 & 37.6 & 4.2 & & & & & 9.3 & 1.0 & & & 0.4 & 0.1 \\
\cline{2-15}
 & 15.0 & 0.8 & 30.3 & 3.4 & & & & & 14.1 & 1.6 & 6.7 & 2.5 & 0.7 & 0.1 
\\
\cline{2-15}
 & 16.6 & 0.7 & 15.6 & 1.8 & & & & & 17.4 & 1.9 & 18.2 & 6.8 & 0.8 & 0.1 
\\
\cline{2-15}
 & 18.2 & 0.7 & 7.4 & 0.8 & & & & & 19.6 & 2.2 & 56.7 & 9.6 & 1.0 & 0.1 
\\
\cline{2-15}
 & 19.6 & 0.6 & 11.9 & 1.4 & & & & & 20.5 & 2.3 & 63.5 & 9.7 & 1.1 & 0.1 
\\
\cline{2-15}
 & 21.0 & 0.6 & 97.4 & 11.0 & & & & & 20.2 & 2.3 & 89.0 & 16.0 & 1.3 & 
0.1 \\
\cline{2-15}
 & 22.6 & 0.5 & 290.8 & 32.6 & & & & & 17.8 & 2.0 & 111.2 & 13.7 & 1.5 & 
0.2 \\
\cline{2-15}
 & 24.5 & 0.5 & 561.0 & 63.0 & 4.9 & 0.9 & & & 18.2 & 2.0 & 132.4 & 16.0 
& 2.7 & 0.3 \\
\cline{2-15}
 & 26.2 & 0.4 & 753.1 & 84.5 & 11.1 & 1.4 & & & 18.3 & 2.0 & 146.1 & 17.6 
& 5.2 & 0.6 \\
\cline{2-15}
 & 27.5 & 0.4 & 852.4 & 95.7 & 16.4 & 2.0 & & & 18.1 & 2.0 & 152.2 & 18.5 
& 7.9 & 0.9 \\
\cline{2-15}
 & 28.7 & 0.3 & 883.8 & 99.2 & 19.3 & 2.4 & & & 17.6 & 2.0 & 144.1 & 17.4 
& 10.5 & 1.2 \\
\cline{2-15}
 & 29.9 & 0.3 & 919.1 & 103.2 & 20.7 & 2.5 & & & 17.7 & 2.0 & 150.8 & 
18.8 & 13.0 & 1.5 \\
\cline{2-15}
 & 31.1 & 0.3 & 933.5 & 104.8 & 20.6 & 2.5 & 0.01 & 0.004 & 17.9 & 2.0 & 
146.7 & 18.2 & 15.8 & 1.8 \\
\cline{2-15}
 & 32.2 & 0.2 & 869.4 & 97.6 & 21.3 & 2.7 & 0.10 & 0.01 & 16.6 & 1.9 & 
158.0 & 19.3 & 17.6 & 2.0 \\
\cline{2-15}
 & 33.3 & 0.2 & 847.1 & 95.1 & 23.1 & 2.9 & 0.38 & 0.04 & 16.7 & 1.9 & 
157.4 & 19.1 & 20.2 & 2.3 \\
\cline{2-15}
 & & &  & &  & &  & &  & &  & &  & \\
\hline
\textbf{CYRIC} & 59.5 & 0.5 & 74.9 & 8.4 &  & & 17.9 & 2.0 & 15.1 & 
1.7 & 122.8 & 13.9 &  & \\
\cline{2-15}
 & 60.3 & 0.5 & 71.9 & 8.2 &  & & 24.9 & 2.8 & 17.5 & 2.0 & 130.4 & 14.7 
&  & \\
\cline{2-15}
 & 61.2 & 0.5 & 76.8 & 8.7 &  & & 31.8 & 3.6 & 16.6 & 1.9 & 121.9 & 13.7 
&  & \\
\cline{2-15}
 & 62.1 & 0.4 & 71.1 & 8.1 &  & & 36.8 & 4.1 & 16.6 & 1.9 & 114.9 & 13.3 
&  & \\
\cline{2-15}
 & 62.9 & 0.4 & 75.7 & 8.6 &  & &  & &  & &  & &  & \\
\cline{2-15}
 & 63.8 & 0.4 &  & &  & & 63.6 & 7.6 & 17.0 & 2.1 & 140.8 & 16.1 &  & 
\\
\cline{2-15}
 & 64.6 & 0.4 & 77.5 & 8.8 &  & & 66.9 & 7.5 & 21.0 & 2.4 & 125.4 & 14.5 
&  & \\
\cline{2-15}
 & 65.5 & 0.3 & 72.5 & 8.1 &  & & 75.9 & 8.5 & 16.4 & 1.9 & 110.5 & 12.8 
&  & \\
\cline{2-15}
 & 66.3 & 0.3 & 71.6 & 8.1 &  & & 84.9 & 9.5 & 17.3 & 2.0 & 127.9 & 14.4 
&  & \\
\cline{2-15}
 & 67.1 & 0.3 & 73.3 & 8.3 &  & & 89.7 & 10.1 & 16.0 & 1.8 & 112.6 & 
13.1 &  & \\
\cline{2-15}
 & 67.9 & 0.3 & 69.3 & 7.9 &  & & 94.7 & 10.6 & 15.3 & 1.7 & 111.6 & 
13.2 &  & \\
\cline{2-15}
 & 68.7 & 0.2 & 63.4 & 7.1 &  & & 121.0 & 13.6 & 19.1 & 2.2 & 145.8 & 
16.9 &  & \\
\cline{2-15}
 & 69.5 & 0.2 & 63.7 & 7.3 &  & & 103.6 & 11.6 & 15.2 & 1.8 & 130.5 & 
15.1 &  & \\
\hline
\end{tabular}

\end{center}
\end{table*}

\begin{table*}[t]
\tiny
\caption{Measured cross sections of the $^{nat}$In(d,x)$^{112m,111g,110g,109}$In and $^{111m,109}$Cd reactions and the estimated mean uncertainties}
\begin{center}
\begin{tabular}{|l|l|l|l|l|l|l|l|l|l|l|l|l|l|l|}
\hline
 & \textbf{Energy} & \textbf{$\Delta$E} & \multicolumn{2}{|c|}{\textbf{$^{112m}$In}} & 
\multicolumn{2}{|c|}{\textbf{$^{111g}$In}} & \multicolumn{2}{|c|}{\textbf{$^{110g}$In}} & \multicolumn{2}{|c|}{\textbf{$^{109
}$In}} & \multicolumn{2}{|c|}{\textbf{$^{111m}$Cd}} & \multicolumn{2}{|c|}{\textbf{$^{109}$Cd}} \\
\cline{2-15}
 &\multicolumn{2}{|c|}{} & \textbf{$\sigma$} & \textbf{$\pm\Delta\sigma$} & \textbf{$\sigma$} & \textbf{$\pm\Delta\sigma$} & 
\textbf{$\sigma$} & \textbf{$\pm\Delta\sigma$} & \textbf{$\sigma$} & \textbf{$\pm\Delta\sigma$} & \textbf{$\sigma$
} & \textbf{$\pm\Delta\sigma$} & \textbf{$\sigma$} & \textbf{$\pm\Delta\sigma$} \\
\cline{4-15}
 & \multicolumn{2}{|c|}{\textbf{MeV}} & \multicolumn{12}{|c|}{\textbf{mbarn}} \\
\hline
\textbf{LLN} & 37.5 & 1.1 & & & 35.2 & 4.0 & 0.7 & 0.1 & & & & & 16.2 
& 2.1 \\
\cline{2-15}
 & 39.5 & 1.0 & & & 31.1 & 3.5 & 1.8 & 0.2 & & & & & 22.3 & 2.7 \\
\cline{2-15}
 & 41.3 & 1.0 & & & 29.2 & 3.3 & 3.4 & 0.4 & & & & & 26.0 & 3.1 \\
\cline{2-15}
 & 43.1 & 0.9 & & & 33.1 & 3.7 & 5.3 & 0.6 & 0.2 & 0.04 & & & 35.6 & 4.1 
\\
\cline{2-15}
 & 44.8 & 0.8 & & & 51.5 & 5.8 & 6.4 & 0.7 & 0.3 & 0.05 & & & 40.6 & 4.7 
\\
\cline{2-15}
 & 46.5 & 0.8 & & & 87.2 & 9.8 & 8.3 & 0.9 & 0.9 & 0.1 & & & 46.4 & 5.3 
\\
\cline{2-15}
 & 48.1 & 0.7 & & & 135.4 & 15.2 & 9.4 & 1.1 & 1.9 & 0.2 & & & 53.7 & 6.1 
\\
\cline{2-15}
 & 49.7 & 0.7 & & & 188.3 & 21.1 & 10.2 & 1.2 & 3.6 & 0.4 & & & 52.3 & 
6.0 \\
\cline{2-15}
 & 51.2 & 0.6 & & & 237.0 & 26.6 & 11.0 & 1.2 & 5.9 & 0.7 & & & 55.5 & 
6.4 \\
\cline{2-15}
 & 52.7 & 0.6 & & & 292.0 & 32.8 & 11.9 & 1.4 & 8.5 & 1.0 & & & 53.8 & 
6.1 \\
\cline{2-15}
 & 54.2 & 0.5 & & & 334.3 & 37.5 & 11.9 & 1.3 & 10.7 & 1.2 & & & 63.7 & 
7.2 \\
\cline{2-15}
 & 55.7 & 0.5 & & & 361.1 & 40.5 & 13.4 & 1.5 & 12.9 & 1.5 & & & 56.5 & 
6.6 \\
\cline{2-15}
 & 57.1 & 0.5 & & & 378.0 & 42.4 & 15.7 & 1.8 & 14.6 & 1.6 & & & 47.3 & 
5.6 \\
\cline{2-15}
 & 58.4 & 0.4 & & & 390.1 & 43.8 & 19.8 & 2.2 & 16.2 & 1.8 & & & 62.0 & 
7.1 \\
\cline{2-15}
 & 59.8 & 0.4 & & & 388.2 & 43.6 & 26.3 & 3.0 & 17.7 & 2.0 & & & 48.4 & 
5.5 \\
\cline{2-15}
 & 61.1 & 0.3 & & & 384.5 & 43.2 & 34.5 & 3.9 & 18.5 & 2.1 & & & 63.6 & 
7.3 \\
\cline{2-15}
 & 62.4 & 0.3 & & & 371.7 & 41.7 & 44.6 & 5.0 & 19.3 & 2.2 & & & 47.8 & 
5.5 \\
\cline{2-15}
 & 63.7 & 0.2 & & & 350.7 & 39.4 & 54.3 & 6.1 & 19.3 & 2.2 & & & 37.3 & 
4.3 \\
\cline{2-15}
 & 65.0 & 0.2 & & & 342.6 & 38.5 & 67.0 & 7.5 & 20.3 & 2.3 & & & 41.9 & 
4.8 \\
\cline{2-15}
 & & &  & &  & &  & &  & &  & &  & \\
\hline
\textbf{VUB} & 8.9 & 1.0 & & & & & & & & & & & & \\
\cline{2-15}
 & 11.2 & 0.9 & & & & & & & & & & & & \\
\cline{2-15}
 & 13.2 & 0.8 & & & & & & & & & & & & \\
\cline{2-15}
 & 15.0 & 0.8 & & & 0.02 & 0.02 & & & & & & & & \\
\cline{2-15}
 & 16.6 & 0.7 & 0.7 & 0.1 & 0.26 & 0.04 & & & & & 0.1 & 0.03 & & \\
\cline{2-15}
 & 18.2 & 0.7 & 2.5 & 0.4 & 1.3 & 0.2 & & & & & 1.2 & 0.1 & & \\
\cline{2-15}
 & 19.6 & 0.6 & 4.0 & 0.5 & 2.3 & 0.3 & & & & & 2.8 & 0.3 & & \\
\cline{2-15}
 & 21.0 & 0.6 & 5.6 & 0.7 & 4.5 & 0.5 & & & & & 4.6 & 0.5 & & \\
\cline{2-15}
 & 22.6 & 0.5 & 5.4 & 0.7 & 5.2 & 0.6 & & & & & 6.2 & 0.7 & & \\
\cline{2-15}
 & 24.5 & 0.5 & 6.9 & 1.0 & 12.2 & 1.4 & & & & & 9.2 & 1.0 & & \\
\cline{2-15}
 & 26.2 & 0.4 & 9.0 & 1.3 & 20.7 & 2.3 & & & & & 11.5 & 1.3 & & \\
\cline{2-15}
 & 27.5 & 0.4 & 8.5 & 1.4 & 29.8 & 3.3 & & & & & 12.5 & 1.4 & & \\
\cline{2-15}
 & 28.7 & 0.3 & 11.7 & 1.9 & 32.8 & 3.7 & & & & & 12.5 & 1.4 & & \\
\cline{2-15}
 & 29.9 & 0.3 & 8.0 & 1.7 & 38.7 & 4.3 & & & & & 11.8 & 1.3 & & \\
\cline{2-15}
 & 31.1 & 0.3 & 16.1 & 3.2 & 41.4 & 4.7 & & & & & 11.1 & 1.2 & & \\
\cline{2-15}
 & 32.2 & 0.2 & 10.7 & 3.7 & 41.4 & 4.6 & & & & & 9.7 & 1.1 & & \\
\cline{2-15}
 & 33.3 & 0.2 & & & 44.6 & 5.0 & & & & & 8.8 & 1.0 & & \\
\cline{2-15}
 & & &  & &  & &  & &  & &  & &  & \\
\hline
\textbf{CYRIC} & 59.5 & 0.5 &  & & 405.5 & 45.5 & 13.9 & 1.6 & 14.5 & 
1.7 &  & &  & \\
\cline{2-15}
 & 60.3 & 0.5 &  & & 404.7 & 45.4 & 18.1 & 2.0 & 12.7 & 1.6 &  & &  & 
\\
\cline{2-15}
 & 61.2 & 0.5 &  & & 433.7 & 48.7 & 20.9 & 2.4 & 12.4 & 1.5 &  & &  & 
\\
\cline{2-15}
 & 62.1 & 0.4 &  & & 404.9 & 45.4 & 23.2 & 2.6 & 16.9 & 2.1 &  & &  & 
\\
\cline{2-15}
 & 62.9 & 0.4 &  & & 422.1 & 47.4 &  & &  & &  & &  & \\
\cline{2-15}
 & 63.8 & 0.4 &  & &  & & 53.0 & 6.1 &  & &  & &  & \\
\cline{2-15}
 & 64.6 & 0.4 &  & & 427.6 & 48.0 & 47.5 & 5.3 & 16.3 & 2.0 &  & &  & 
\\
\cline{2-15}
 & 65.5 & 0.3 &  & & 400.0 & 44.9 & 55.6 & 6.3 &  & &  & &  & \\
\cline{2-15}
 & 66.3 & 0.3 &  & & 383.7 & 43.1 & 66.8 & 7.5 & 19.5 & 2.2 &  & &  & 
\\
\cline{2-15}
 & 67.1 & 0.3 &  & & 379.2 & 42.6 & 66.3 & 7.4 & 19.1 & 2.2 &  & &  & 
\\
\cline{2-15}
 & 67.9 & 0.3 &  & & 361.2 & 40.5 & 72.3 & 8.1 & 15.5 & 1.9 &  & &  & 
\\
\cline{2-15}
 & 68.7 & 0.2 &  & & 330.4 & 37.1 & 100.3 & 11.3 & 34.8 & 4.2 &  & &  
& \\
\cline{2-15}
 & 69.5 & 0.2 &  & & 331.3 & 37.2 & 88.9 & 10.0 &  & &  & &  & \\
\hline
\end{tabular}

\end{center}
\end{table*}

\subsection{Integral yields}
\label{3.2}
From excitation functions obtained by a spline fit to our experimental cross section data integral thick target yields were calculated and are shown in Fig. 14 as a function of the energy for the medically relevant $^{113}$Sn, $^{111}$Sn, $^{110}$Sn, $^{114m}$In and $^{111g}$In, radionuclides. Due to the favorable production and decay characteristics these radioisotopes are also of importance in Thin Layer Activation technique (TLA) \citep{TLA, Warner}. For $^{113g}$Sn an acceptable agreement with the results of \citep{Dmitriev1975} is seen.

\begin{figure}
\includegraphics[width=0.5\textwidth]{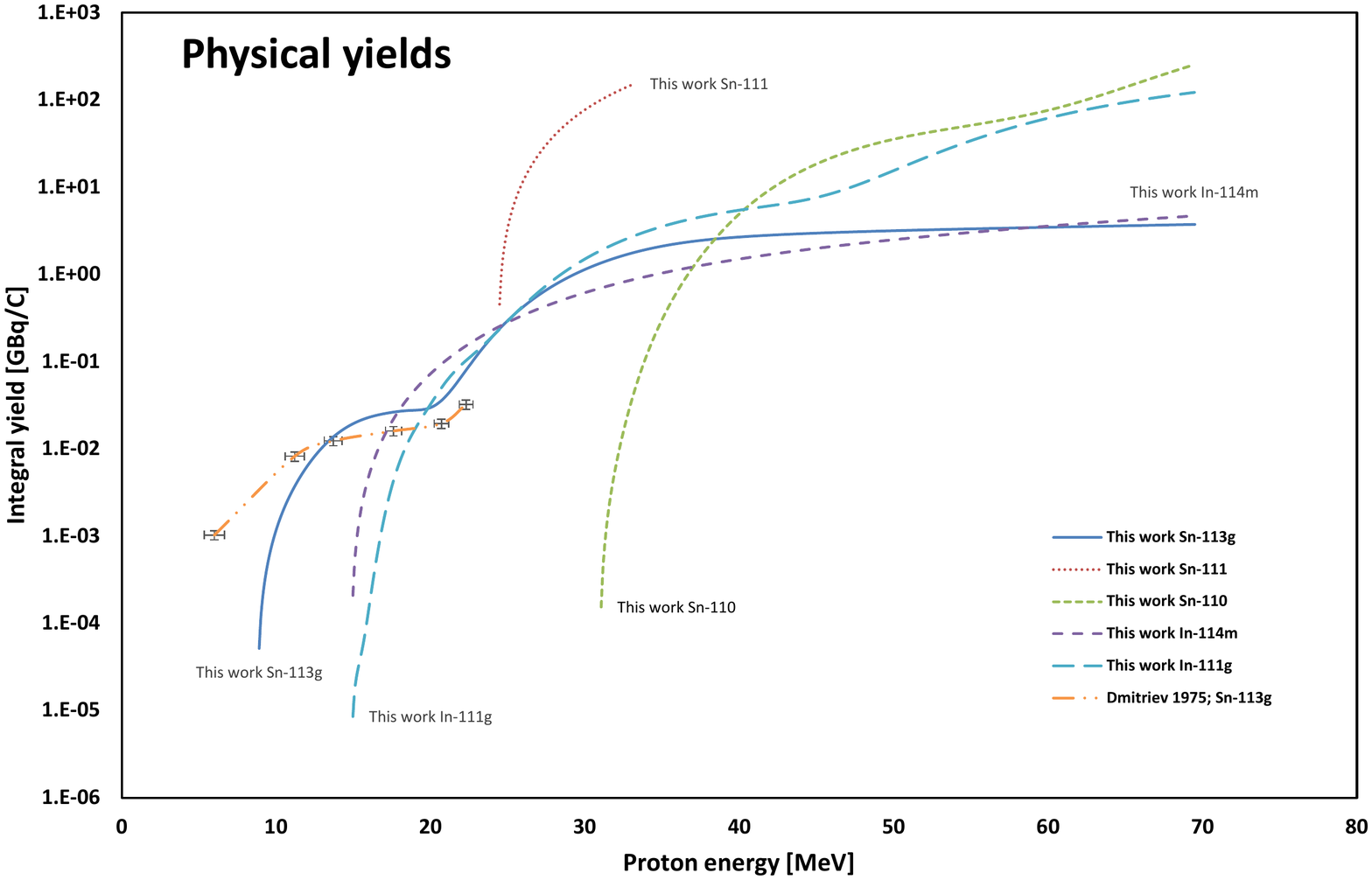}
\caption{Integral yields of the 110, 111,$^{113g}$Sn and 111g,$^{114m}$In calculated from the measured excitation functions (this work) compared with the literature data}
\label{fig:14}       
\end{figure}

\section{Applications}
\label{4}

\subsection{Medical applications}
\label{4.1}
Various radiopharmaceuticals labeled with $^{111g}$In (T$_{1/2}$= 2.81 d, 100 \% EC decay) are used in the diagnosis of cancer and other diseases through SPECT. However, in particular for receptor-type studies, the quantification of the uptake of the radiopharmaceutical via PET measurements may be important. For $^{111g}$In the corresponding isotope of choice is the $\beta^+$-emitter $^{110m}$In (61.3 \%). The Auger- electron emitter $^{113m}$In is a candidate for internal radio-therapy, while recently the favorable Auger-electron characteristics of $^{111g}$In and $^{114m}$In were recognized and are used for studies of internal radio-immunotherapy.

\subsubsection{$^{114m}$In}
\label{4.1.1}
Production routes for $^{114m}$In through proton and deuteron induced nuclear reactions on cadmium, indium and tin were measured and reviewed by us \citep{TF2005,  TF2011, TF2006, TF2007}. It was shown there that $^{114m}$In can be produced also by using high energy reactions on In with high yield. However the product is not carrier free, which limits its possible applications.

\subsubsection{$^{113m}$In}
\label{4.1.2}
Comparison of production routes of $^{113m}$In and its $^{113}$Sn parent were discussed in detail in our previous work on deuteron induced reactions on indium. \citep{TF2011}. By investigating all these reactions in detail it can be concluded that by direct production it is difficult to produce $^{113}$In with high radionuclidic purity. Large scale production of $^{113}$Sn at present is done via the $^{112}$Sn(n,$\gamma$) reaction. This route results in a product of low specific activity and asks for highly enriched targets. Promising production routes, utilizing charged particle beams to reach high specific activities, can be proposed relying on the $^{111}$Cd($\alpha$,2n), $^{nat}$Cd($\alpha$,xn), $^{nat}$Cd($^{3}$He,xn), $^{113}$In(p,n), $^{nat}$In(p,xn) and $^{113}$In(d,2n) nuclear reactions. By intercomparing these production routes at medium energies the $^{113}$In(d,2n) and $^{nat}$In(p,xn) reactions are the most productive (Table 4). At even higher energies (Ep $\>=$ 100 MeV) the $^{nat}$In(d,xn) reaction is also advantageous.

\begin{table*}[t]
\tiny
\caption{Comparison of charged particle production routes of $^{113}$Sn on indium}
\begin{center}
\begin{tabular}{|l|l|l|l|l|}
\hline
\textbf{Reaction} & \multicolumn{2}{|c|}{\textbf{Cross section at energy}} & \textbf{Energy range} & \textbf{Activity} \\
\cline{2-5}
 & \textbf{mb} & \textbf{MeV} & \textbf{MeV} & \textbf{GBq} \\
\hline
$^{113}$In(p,x)$^{113}$Sn & 650 & 11 & 20-5 & 22.7 \\
\hline
$^{nat}$In(p,x)$^{113}$Sn & 67 & 11.5 & 18-6 & 2.1 \\
\cline{2-5}
 & & & 40-6 & 10.0 \\
\hline
$^{113}$In(d,x)$^{113}$Sn & 970 & 15.5 & 40-5 & 71.9 \\
\hline
$^{nat}$In(d,x)$^{113}$Sn & 41.5 & 15 & 20-8 & 1.3 \\
\cline{2-5}
 & & & 40-8 & 42.4 \\
\hline
\end{tabular}

\end{center}
\end{table*}

\subsubsection{$^{111g}$In}
\label{4.1.3}
We made detailed measurements and reviews on the production routes of the widely used $^{111g}$In \citep{Hermanne2014b, TF2005, TF2006}. A review of $^{111g}$In production was recently done \citep{Lahiri}. To avoid radionuclide impurities and ensure high specific activity in practice the $^{111}$Cd(p,n) and $^{112}$Cd(p,2n) are used, asking for enriched targets and low to medium energy. The yields and radionuclide purity are high, but highly enriched targets are required to avoid the $^{114m}$In long-lived side product. 
In principle the $^{111}$Cd(d,2n) reaction could be used as the production yield is slightly higher (if high enough deuteron energy is available) than what is obtained by using protons. In practice, however, the deuteron induced reaction cannot compete with the proton induced reaction for routine production of $^{111g}$In as only a very limited number of accelerators can provide the required higher deuteron beams.
The conclusion is nearly the same for use of the $^{109}$Ag($\alpha$,2n)$^{111}$In reaction, where natural targets can be used and no $^{114m}$In side product is formed. The negative factors are the lower yield and the very limited availability of alpha beams above 20 MeV. 
By using indium target the carrier free products can be reaching only through indirect methods. The cross section of the $^{nat}$In(p,x)$^{111}$Sn-$^{111g}$In process is high in the 20-50 MeV energy range (200-300 mb). The main disadvantage is the small batch yield due to the short half-life of the parent $^{111}$Sn (short irradiation time, losses during the separation).

\subsubsection{$^{110m}$In}
\label{4.1.4}
Comparison of production routes of $^{110m}$In and its parent $^{110}$Sn were also done in recent work dedicated to this purpose \citep{TF2015}.  The $^{110}$Sn($^{110m}$In)  generator could be prepared at low and medium energies with light charged particles via the $^{113}$In(p,4n), $^{nat}$In(p,xn), $^{110}$Cd($\alpha$,4n), $^{nat}$Cd($\alpha$,xn), $^{110}$Cd($^{3}$He,3n), $^{nat}$Cd($^{3}$He,xn)  reactions. Each of these routes require a proper selection of an adapted energy range and high incident energies. The main advantage of the indirect method is the high radionuclide purity, which can be easily assured by the proper irradiation and cooling parameters. Considering the available commercial accelerators, use of the $^{nat}$In(p,xn)$^{110}$Sn reaction seems to be the simplest and most productive method, however it requires 70-100 MeV accelerators. It should be mentioned that the generator can be produced also at lower energy machines, but it requires highly enriched targets \citep{TF2015}.
For direct production the $^{110}$Cd(p,n), $^{110}$Cd(d,2n),  $^{107}$Ag($\alpha$,n) and $^{109}$Ag($^{3}$He,2n) reactions mean the candidate routes. Lower energy accelerators can be used, but highly enriched targets are required. 

\subsection{Thin layer activation (TLA)}
\label{4.2}
Among the investigated reaction products, the production and decay parameters of $^{113}$Sn (115.09 d, 391.698 keV - 64.97 \%) are satisfying the usual requirements (medium half-lives, intense gamma-lines) for application in TLA.
Recommended cross sections and specific activity as a function of depth up to 20 MeV can be found in the IAEA TLA database for the $^{nat}$In(p,x)$^{113}$Sn reaction. According to Fig. 15 the recommended data are a little lower above 12 MeV and a little higher below 12 MeV proton energy, than what was found in this experiment.  
By studying the decay parameters of the reaction products and their excitation functions it can be concluded that the $^{113}$Sn can be used in TLA not only in low energy irradiation but much more effectively at higher energy irradiations due to the significantly higher cross sections, i.e. shorter irradiation time. (Fig. 2). It should also be mentioned that in this case the penetration depth will be much higher, which might not fit to every wear measurement task.
It should be mentioned additionally to $^{113}$Sn that the $^{nat}$In(p,x)$^{114m}$In reaction is also a good candidate for TLA labeling, due to the  shorter half-life, high cross sections and the  proper decay  of $^{114m}$In ( 49.51 d, 190.27 keV - 15.56 \%). According to Fig. 6, however, unfortunately only one experimental data set is available.

\begin{figure}
\includegraphics[width=0.5\textwidth]{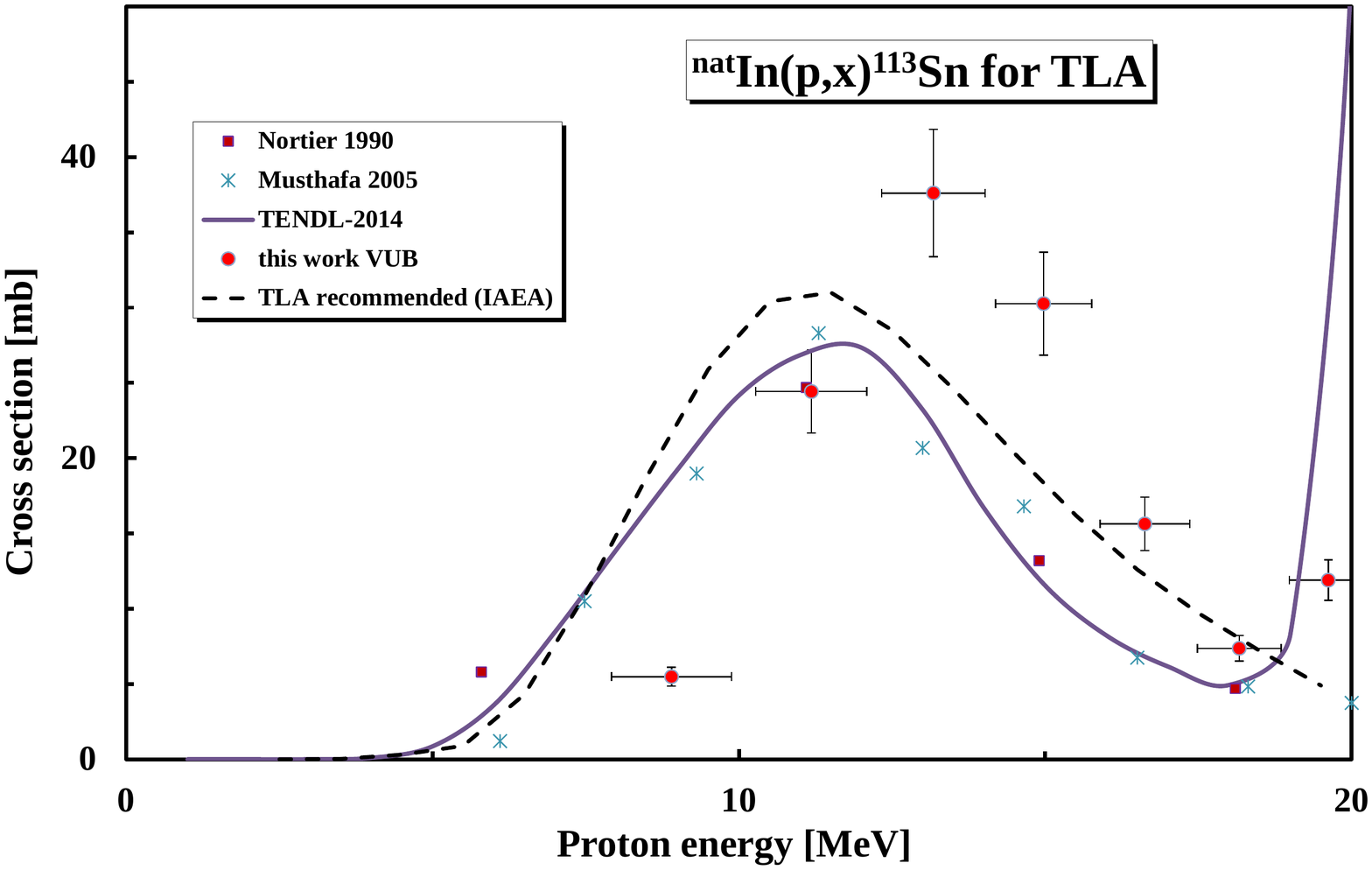}
\caption{Experimental cross section of $^{nat}$In(p,x)$^{113}$Sn reaction in comparison with the recommended data from the IAEA TLA database}
\label{fig:15}       
\end{figure}

\section{Summary and conclusions}
\label{5}
In this study excitation functions for production of the radioisotopes  $^{113,111,110}$Sn, $^{115m,114m,113m,112m,111,110g}$In and $^{111m,109}$Cd were determined up to 70 MeV proton energy on indium targets, out of them $^{115m,114m,113m,112m,111,110g}$In and $^{111m}$Cd  for the first time. The results might contribute to medical and industrial applications as well as to development of theoretical nuclear reaction codes.
We discuss the usefulness of proton induced reactions on indium for production of $^{110}$Sn/$^{110m}$In, $^{111}$Sn/$^{111g}$In, $^{113}$Sn/$^{113m}$In, and $^{114m}$In in comparison with other production routes. Except $^{114m}$In, where the product is carrier added, the advantage for the production of other three products through Sn/In generators is the high radionuclide purity and high specific activity. Disadvantage of the generator method in case of $^{111g}$In is the short half-life of $^{111}$Sn (small batch yields). 
The recently measured new experimental data will also allow further development of the nuclear database for thin layer activation, as well as make possible a further refinement in the nuclear reaction model codes.

\section{Acknowledgements}
\label{}
This work was done in the frame of MTA-FWO (Vlaanderen) research projects. The authors acknowledge the support of research projects and of their respective institutions in providing the materials and the facilities for this work.
 



\bibliographystyle{elsarticle-harv}
\bibliography{Inp}







\end{document}